# Decentralized Collaborative Knowledge Management using Git

Natanael Arndt[a,*], Patrick Naumann[b], Norman Radtke[a], Michael Martin[a], Edgard Marx[a,b]

[a]*Agile Knowledge Engineering and Semantic Web (AKSW), Institute of Computer Science, Leipzig University, Augustusplatz 10, 04109 Leipzig, Germany*
[b]*Hochschule für Technik, Wirtschaft und Kultur Leipzig (HTWK), Gustav-Freytag-Str. 42A, 04277 Leipzig, Germany*

**Abstract**

The World Wide Web and the Semantic Web are designed as a network of distributed services and datasets. The distributed character of the Web brings manifold collaborative possibilities to interchange data. The commonly adopted collaborative solutions for RDF data are centralized (e. g. SPARQL endpoints and wiki systems). But to support distributed collaboration, a system is needed, that *supports divergence* of datasets, brings the possibility to *conflate diverged states*, and allows distributed datasets to be *synchronized*. In this paper, we present *Quit Store*, it was inspired by and it builds upon the successful Git system. The approach is based on a formal expression of evolution and consolidation of distributed datasets. During the collaborative curation process, the system automatically versions the RDF dataset and tracks provenance information. It also provides support to branch, merge, and synchronize distributed RDF datasets. The merging process is guarded by specific merge strategies for RDF data. Finally, we use our reference implementation to show overall good performance and demonstrate the practical usability of the system.

## 1. Introduction

Apart from documents, datasets are gaining more attention on the World Wide Web. An increasing number of the datasets on the Web are available as Linked Data, also called the *Linked Open Data Cloud*[1] or *Giant Global Graph*[2]. Collaboration of people and machines is a major aspect of the World Wide Web and as well of the Semantic Web. Currently, the access to RDF data on the Semantic Web is possible by applying the Linked Data principles [9], and the SPARQL specification [41], which enables clients to access and retrieve data stored and published via SPARQL endpoints. RDF resources in the Semantic Web are interconnected and often correspond to previously created vocabularies and patterns. This way of reusing existing knowledge facilitates the modeling and representation of information and may optimally reduce the development costs of a knowledge base. However, reusing existing RDF resources (terminological as well as instance resources) causes problems in locating, applying, and managing them. The administrative burden of these resources then increases immensely, insofar as the original sources change (partially) and the reuse of these RDF resources takes place collaboratively and in a decentralized manner. For example, this type of reuse occurs during the creation of a domain-specific vocabulary or a specific set of instances developed by organizationally independent collaborators. Presumably, collaboration in such a setup is either primarily agile or it can be organized top-down, in which case it has to be completely supervised. However, this complete supervision requires a high amount of effort. As a result, structural and content interferences as well as varying models and contradictory statements are inevitable.

Projects from a number of domains are striving for distributed models to collaborate on common knowledge bases. In the domain of e-humanities, the projects Pfarrerbuch[3], Catalogus Professorum[4] [40], Héloïse – European Network on Digital Academic History[5] [39], and Professorial Career Patterns of the Early Modern History[6] are good examples of the need to explore and track provenance and the evolution of the domain data. In the context of managing historical prosopographical data, the source of the statements is relevant to evaluate their credibility and to consider the influence of their environment. In libraries, metadata of electronic library resources are gathered and shared among stakeholders. The AMSL[7] project aims to collaboratively curate and manage electronic library resources as Linked Data [4, 38]. In a collaborative data curation setup we need to identify the origin of any statement introduced into a dataset. This is essential in order to be able to track back the conclusion of license contracts and identify sources of defective metadata. But even enterprises have a need to manage data in distributed setups. The LUCID – Linked Value Chain Data[8] project [20] researches the communication of data along supply chains The LEDS – Linked Enterprise Data Services[9] project focuses on how to organize and support distributed collaboration on datasets for the management of background knowledge and business procedures.

*Corresponding author
*Email addresses:* `arndt@informatik.uni-leipzig.de` (Natanael Arndt), `pnaumann86@gmail.com` (Patrick Naumann), `radtke@informatik.uni-leipzig.de` (Norman Radtke), `martin@informatik.uni-leipzig.de` (Michael Martin), `marx@informatik.uni-leipzig.de` (Edgard Marx)
*URL:* `http://aksw.org/NatanaelArndt` (Natanael Arndt), `http://aksw.org/NormanRadtke` (Norman Radtke), `http://aksw.org/MichaelMartin` (Michael Martin), `http://aksw.org/EdgardMarx` (Edgard Marx)

[1] `http://lod-cloud.net/`
[2] `http://dig.csail.mit.edu/breadcrumbs/node/215`
[3] `http://aksw.org/Projects/Pfarrerbuch`
[4] `http://aksw.org/Projects/CatalogusProfessorum`
[5] `http://heloisenetwork.eu/`
[6] `http://catalogus-professorum.org/projects/pcp-on-web/`

Currently, the collaboration on Linked Data Sets is mainly done by keeping a central version of a dataset. In this context, collaborators edit the same version of the dataset simultaneously. The systems available enable collaboration on Linked Data are central SPARQL endpoints and wiki systems [19, 18, 33]. In both cases, a common version of a dataset is kept in a central infrastructure and thus collaboration happens on a single, shared instance. This central approach for a synchronized state has drawbacks in scenarios in which the existence of different versions of the dataset is preferable. Furthermore, the evolution of a dataset in a distributed setup is not necessarily happening in a linear manner. Multiple versions of a dataset occur if the participants do not all have simultaneous access to the central dataset (for instance, if they are working from mobile devices with limited network connection). If a consensus on the statements in a dataset hat not yet been reached, multiple viewpoints need to be expressed as different versions of the dataset. Hence, a system that fosters the evolution of a dataset in a distributed collaborative setup needs to

- **support divergence** of datasets;
- **conflate diverged states** of datasets; and
- **synchronize** different distributed derivatives of the respective dataset.

As a consequence of conflating diverged datasets, the utilized system also needs to

- identify possible occurring conflicts and contradictions, and
- offer workflows to resolve identified conflicts and contradictions.

In the early days of computers, the term *software crisis* was coined to describe the immaturity of the software engineering process and software engineering domain. Dijkstra described the situation as follows:

> [...] as long as there were no machines, programming was no problem at all; when we had a few weak computers, programming became a mild problem, and now we have gigantic computers, programming has become an equally gigantic problem.[10]

The process of creating software could be made more reliable and controllable by introducing software engineering methods. In the 1970s, software configuration management enabled structured collaborative processes, where version control is an important aspect for organizing the evolution of software. Early *version control systems* (VCS), such as *CVS* and *Subversion*, allowed central repositories to be created. The latest version on the repository represents the current state of development, and the linear versioning history draws the evolution process of the software. Distributed VCS (DVCS), such as *Darcs*, *Mercurial*, and *Git*, were developed to allow every member of a distributed team to fork the current state of the programs source code and individually contribute new features or bug-fixes as pull-requests, which then can be merged into a master branch, that represents the current stable version.

Learning from software engineering history where DVCS has helped to overcome the software crisis, we claim that adapting DVCS to Linked Data is a means to support decentralized and distributed collaboration processes in knowledge management. The subject of collaboration in the context of Linked Data are datasets instead of source code files, *central* VCS systems correspond to *central* SPARQL endpoints and wiki systems. Similar to source code development with DVCS, individual local versions of a dataset are curated by data scientists and domain experts. Tracking provenance during the process of changing data is a basic requirement for any version control system. It is therefore important to record the provenance of data at any step of a process, that involves possible changes of a dataset (e. g., creation, curation, and linking).

Our aim is to provide a system that enables distributed collaboration of data scientists and domain experts on RDF datasets. In particular we focus on a *generic* solution that allows us to target the problem of collaboration. By *generic*, we mean that the solution should not make any assumptions on the application domain. This includes the fact that the system has to rely on the pure RDF data model and not rely on or add support for additional semantics such as OWL or SKOS. On the informal *Semantic Web Layer Cake* model[11], we focus on the syntactic *data interchange* layer in combination with the *generic* SPARQL query language. To support distributed collaboration on the *data interchange* layer, we propose a methodology of using a Git repository to store the data in combination with a SPARQL 1.1 interface to access it. We introduce the *Quit Stack* ("Quads in Git") as an integration layer to make the collaboration features of Git repositories accessible to applications operating on RDF datasets.

[7] `http://amsl.technology/`
[8] `http://www.lucid-project.org/`
[9] `http://www.leds-projekt.de/`
[10] `https://www.cs.utexas.edu/users/EWD/transcriptions/EWD03xx/EWD340.html`
[11] `https://www.w3.org/2007/03/layerCake.svg`

In this paper, we have combined multiple aspects for supporting the collaboration on RDF datasets. Regarding the individual improvements to these aspects, which were discussed independently, we can now present a combined and comprehensive system. For this we first introduce a formal model to describe and transform the operations implemented in Git to operations on RDF datasets. The formal model was published in [2]. In this paper, we can present a more elaborated formal model that was especially extended and improved regarding the atomic partitioning. The formal model is used to express changes and describe the evolution of datasets with support for tracking, reverting, branching, and merging. This model provides support for named graphs and blank nodes. To actually pursue merging operations (i. e. identifying and resolving conflicts), we propose various merge strategies that can be utilized in different scenarios. Initial functionality for tracking and analyzing data-provenance using the *Quit Store* was examined in [3]. Since this paper, we have improved and clarified the handling of the provenance data and can present an extended data model. We enable access to provenance-related metadata retrieved from RDF data that is managed in a Git repository. It further supports collaborative processes by providing provenance mechanisms to track down and debug the sources of errors. To present and further examine the possibilities to support distributed collaborative dataset curation processes, we have created a reference implementation. We have published the description of the initial prototype of the *Quit Store* in [5]. In the current paper, we can present an improved implementation that provides support for provenance tracking and exploitation. Furthermore we were able to increase the overall performance of the system by implementing an overworked architecture, which is mainly covered in section 9. The new system also has a ready-to-use interface to create, query, and merge different branches, where the user can select between multiple merge strategies.

The paper is structured as follows. We present the description of an application domain with relevant use cases that are presented in section 2. Requirements for a decentralized collaboration setup are formulated in section 3. The state of the art is discussed in section 4 followed by relevant preliminaries in section 5, such as an introduction to Git and a discussion of RDF serialization and blank nodes. An introduction to the preliminaries of the formal model with basic definitions are given in section 6. Based on the definitions in section 6, the basic operations on a versioning graph of distributed evolving datasets are defined in section 7. As an extension of the basic operations, multiple merge strategies are presented in section 8. The approach and methodology of the system as well as the prototypical reference implementation is specified in detail in section 9. The concepts presented are evaluated regarding correctness and performance, using our prototypical implementation, in section 10. Finally, we discuss the results of the paper in section 11 and a conclusion is given together with a prospect for future work in section 12.

Throughout the paper, we use the following RDF-prefix mappings:

```
@prefix quit: <http://quit.aksw.org/vocab/> .
@prefix prov: <http://www.w3.org/ns/prov#> .
@prefix foaf: <http://xmlns.com/foaf/0.1/> .
@prefix local: <http://quit.local/> .
@prefix ex: <http://example.org/> .
@prefix rdf: <http://www.w3.org/1999/02/22-rdf-syntax-ns#> .
@prefix rdfs: <http://www.w3.org/2000/01/rdf-schema#> .
@prefix xsd: <http://www.w3.org/2001/XMLSchema#> .
```

In most cases, we could reuse existing terms from the `prov:` and `foaf:` vocabularies to express the provenance information produced by the quit store. The `quit:` vocabulary was created to define additional terms and instances that were not defined in existing vocabularies. Furthermore, it also provides terms to express the software configuration of the *Quit Stack*. We use the namespace `local:` to denote terms and instances that are only valid in a local context of a system. The namespace `ex:` is used for exemplary instances that are placeholders for instances whose IRIs depend on the exact application domain.

## 2. Domain and Derived Use Cases

The research group Agile Knowledge Engineering and Semantic Web (AKSW) together with historians from the Herzog August Library in Wolfenbüttel (HAB) run a research project in cooperation with partners from the Working Group of the German Professor's catalogs and the European research network for academic history–Héloïse. The project's aim is to develop a new research method across humanities and the computer science field employing Semantic Web and distributed online databases for study and evaluation of collected information on groups of historical persons focusing on German professors' career patterns during the 18th and 19th centuries. The project deals with many challenges as individual datasets have been developed by those different communities, some of them are more than 600 years old as the *Catalogus Professorum Lipsiensium*. These communities have different singularities that make the development and management of a common vocabulary very challenging. For instance, the German professor's dataset of the state of Saxony contains a list of as much as six working places and respective positions of each professor across time as well as a detailed description of the archive where the information were extracted from. These information were not available in previous published datasets and therefore they would cause a vocabulary change. However, in order to perform the changes, there is a need of (a) creating an independent vocabulary or (b) to implement the changes over a mutual agreement among the different research groups. Creating independent vocabularies (a) in the individual working groups would not help to integrate the datasets. Due to the organizational distribution of the individual research groups, a common vocabulary expressing a mutual agreement among the groups (b) can only be the result of a collaborative process. Based on the origin of the individual datasets, their diverse granularity and structure the individual discussions have varying starting points.

This description of an application domain is exemplary for other interdisciplinary projects in fields as the digital humanities, but can also stand for many other collaborative scenarios. Our aim is to provide a system to support the distributed collaboration of data scientists and domain experts on RDF datasets. To break this domain description down we have identified several use cases that can contribute to a comprehensive system to support such collaborative evolutionary processes. In the following we describe the use cases:

- UC 1: *Collaborative and Decentralized Design*
- UC 2: *Crowdsourcing Information with Citizen Scientists*
- UC 3: *Integrating With Existing Systems*
- UC 4: *Backup*
- UC 5: *Provenance Recording and Exploration*

UC1 *Collaborative and Decentralized Design.* Vocabularies and datasets are used to encode a common understanding of describing information that represent and express a given domain from the point of view of the creators at a certain time. Until a formulation of the common understanding or definition of a term or dataset item is reached different proposals are discussed in parallel. The process of creating a vocabulary or a dataset is long and involves many aspects such as the targeted users, the domain, used technology, and the maintenance of the vocabulary over time. It can be collaborative, agile and usually implicates many iterations until it reaches a mature state. For instance the creation and evolution of the `foaf:` vocabulary went through 10 public iterations of the specification from 2005 until 2014[12]. In a more generalized way the distributed collaboration allows users to work together through individual contributions. It played an important roll on the evolution of the World Wide Web. Good examples of collaborative systems are Wikipedia[13] respective Wikidata[14]. The Wikipedia project provides a platform for collaboration and exchange that allows volunteers around the world to collectively curate its content in a crowd-sourcing manner. However, users might have different needs and requirements. Heterogeneous distributed collaboration systems can support a multitude of data perspectives and a decentralized fashion allows users to evolve their dataset copy distinctly while still sharing a common portions of it.

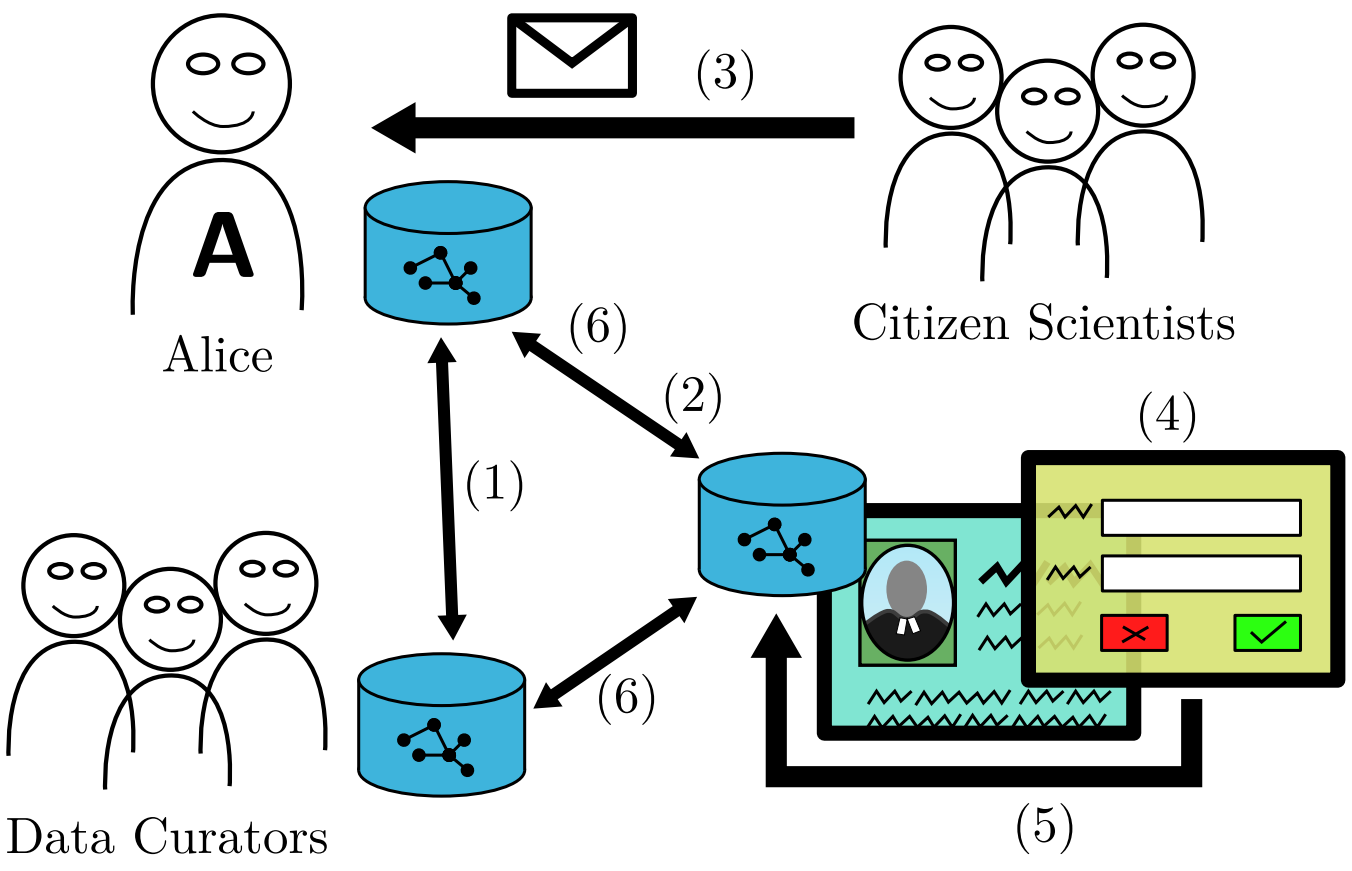

Figure 1: Integrating a crowdsourcing process into a collaborative curation process.

UC2 *Crowdsourcing Information with Citizen Scientists.* Collaborative processes can also be used to incorporate external non-professional collaborators into a research workflow. This should be demonstrated with the following prototypical user story. Alice collaborates with a team on some scientific set of open data (fig. 1, 1). Alice has published the data together with an exploration interface on the Web (fig. 1, 2). While Alice's team of curators works hard on the data there are some citizen scientists who send e-mails to Alice reporting errors in the data (fig. 1, 3). Since Alice and her team work hard there is no time to read through all the prose of the citizen scientists and then manually seek the property in the dataset that has to be corrected. To incorporate the contributions of the citizen scientists into the existing collaboration workflow, Alice adds a data editor to the exploration interface on the Web (fig. 1, 4). The exploration interface can then be used by the citizen scientists to fix errors, they have found. All changes made by the citizen scientists are then transferred to the collaboration repository (fig. 1, 5) that is synchronized with the repositories of the team of curators (fig. 1, 6). Now each member of the team of curators can review the proposed changes and incorporate them into the dataset.

UC3 *Integrating With Existing Systems.* Tools to create and edit RDF knowledge bases exist, but might lack support for collaborative scenarios. It would involve a high complexity to attach existing single place editing and exploration systems with collaboration functionality. A possibility to attach single place systems to a collaboration infrastructure by using well defined interfaces would allow the creation of collaborative setups without the need of new tools. An architecture that employs well defined interfaces for existing tools would also follow and support the single responsibility principle as formulated in the context of the "UNIX philosophy" as *Make each program do one thing well* [22, 23]. Users can thus continue using their existing tools, daily workflows, and familiar interfaces.

[12] `http://xmlns.com/foaf/spec/`
[13] `https://www.wikipedia.org/`
[14] `https://www.wikidata.org/`

UC4 *Backup.* Data storage systems can be attacked, hacked, and fail, data can be corrupted. Some users might desire to prevent data loss by periodically performing dataset backups for instance by synchronizing them with a remote system. By doing so, it is easier to restore it whenever necessary. RDF is not different than any other data, it is always important to create backups. A version controlled backup system also allows to restore data even after faulty changes. Further a backup of the work of each collaborator allows other parties to continue the overall work, even if collaborating parties quit. Providing means to backup data, helps to avoid content loss as well as time on restoring it. By integrating a tool to track the changes of the data and submit the data to a safe location into the daily work of the data creator, avoids gaps in the backup and distraction from the main tasks.

UC5 *Provenance Recording and Exploration.* When integrating datasets from different sources or performing update operations, recording of provenance information can be desired. According to the Oxford online dictionary, *provenance* has to be with the *"origin or earliest known history of something"*[15]. The storage of data's provenance all the way down to the atomic level (insertions and deletions) can be very useful to backtrack the data transformation process and spotlight possible errors on different levels during the data life-cycle [8] as well as in the data itself. Therefore, the provenance can be explored to improve data management and engineering processes. To support developers of provenance systems, Groth et al. [25] provide general aspects that should be considered by any system that deals with provenance information. The aspects are categorized as follows [25]:

- *Content* describes what should be contained in provenance data, whereas entities, contributing sources, processes generating artifacts, versioning, justification, and entailment are relevant dimensions.

- *Management* refers to concerns about how provenance should be captured and maintained, including publication and access, dissemination, and how a system scales.

- *Use* is about how user specific problems can be solved using recorded provenance. Mentioned are understanding, interoperability, comparison, accountability, trust, imperfections, and debugging.

## 3. Requirements

Based on our use cases we formulate the following requirements for a distributed collaboration system on RDF datasets. The requirements point out aspects of supporting a distributed setup of data curators in collaborating on RDF datasets as well as aspects of the system that are needed to record and explore the evolution of the data. Requirements for collaborative vocabulary development are already formulated in [27]. We adopt some of these requirements that overlap with our use cases. In contrast to [27] we focus on the technical collaboration on a distributed network, rather than the specific process of creating vocabularies. First we present the three major requirements that are directly implied by our aim to support distributed collaboration (REQ 1 to 3). The major requirements are followed by five requirements that are needed to support the distributed collaboration process (REQ 4 to 6) respective are necessary to employ the system in a semantic web context (REQ 7 and 8).

REQ1 *Support Divergence.* In collaborative scenarios contributors may differ in their motivation to contribute, for example because of their organizational role, may differ in their opinion to the subject of collaboration, or may provide contradicting contributions in any other way. Thus the existence of multiple different versions of the dataset is preferable, for instance to express dissensus, disagreement, or situations in which a consensus was not yet reached. Especially in distributed collaborative setups (cf. UC 1 and 2) it can happen, that collaborating parties are not always sharing a common understanding of a certain topic. But also because of organizational structures the evolution of a dataset is not necessarily happening in a linear manner, for instance if partial subjects are discussed in sub-working groups. Thus the system needs to be able to handle diverging states of a dataset.

REQ2 *Conflate Diverged States.* The aim of collaboration is to eventually contribute to a common subject. To combine the contributions in diverged states of a dataset to a common dataset a possibility to conflate the diverged states is needed. Because diverged states can encode dissensus it is necessary to identify possible conflicts before actually merging the datasets. The definition of conflict depends on the semantics of the data and the application domain, thus different strategies are needed to identify conflicts. When possibly identifying conflicts, the system needs to offer workflows to resolve identified conflicts and contradictions.

REQ3 *Synchronize.* Collaborating on a single centralized copy of a dataset causes much coordination difficulties. For instance if a simultaneous access to the central dataset is not possible for all participants (e.g. from mobile devices). Thus we focus on a distributed setup of collaboration systems. To support the collaboration of multiple distributed parties across system boundaries (cf. UC 1 and 2) it is necessary to exchange data between systems. The system should support the synchronization of their available states of the dataset with the available states of remote setups. This is also important to keep up a collaborative process if participating systems fail (cf. UC 4). This synchronization process can happen in real-time or asynchronously.

[15] `https://en.oxforddictionaries.com/definition/provenance`

REQ4 *Provenance of Contributions.* Provenance information have to be attached to a contribution to the common dataset, at least a change reason, author information, and date of commit. For the automated interaction with the system, executed operations have to be documented, e. g. the query or data source. To utilize the provenance information that are recorded during the evolution of a dataset, they need to be represented in a queriable graph. Access to this graph has to be provided through a query interface. The interface has to return a structured representation of metadata recorded for the selected versions of a dataset. This is necessary as a prerequisite of the aspects publication and access, resp. dissemination of the provenance information (cf. UC 5: *Management*; *Communication support (R1)* and *Provenance of information (R2)* in [27])

REQ5 *Random Access to any Version.* For a collaboration system it should be possible to randomly access any version of the dataset without the need for a rollback of the latest versions resp. resetting the storage system. For instance, when collaborators are currently working on different versions of the data (cf. UC 1). This allows queries across versions and it can support the debugging process (*Use*, cf. UC 5). To restore any time slice of a backup it is also necessary to access and restore an arbitrary version (cf. UC 4).

REQ6 *Deltas Among Versions.* It is required to calculate the difference between versions generated by contribution of collaborators. The difference should express the actual changes on the dataset rather than changes to the serialization of the data in the repository. This is necessary when reviewing external contributions to a dataset (cf. UC 2). It is also a prerequisite to explore the provenance information, for the debug operation (*Use*, cf. UC 5), and to analyze backups of datasets (cf. UC 4). The calculated difference should be expressed in a machine readable format. (cf. *Deltas among versions (R7)* in [27])

REQ7 *Support of RDF Data Sets and Modularization of Graphs.* The system should be able to handle multiple RDF graphs (i.e. RDF datasets), in a repository. This allows users resp. collaborators to organize the stored knowledge in individual organizational units, as it is required by their application. This requirement also provides the functionality to implement the requirement *Modularity (R9)* as formulated in [27]. The method works with different granularities of modularization of RDF datasets. This is of interest when the system should be integrated with existing systems (cf. UC 3).

REQ8 *Standard Data Access Interface.* Different implementations of collaboration interfaces can access and collaborate on a common repository (cf. UC 3). Collaborators can use different RDF editors to contribute to the repository (cf. UC 1). To some extent the methodology should even be robust to manual editing of RDF files contained in the repository. In contrast to the requirement *Editor agnostic (R8)* as formulated in [27], we do not require the syntax independence on the repository and understand the editor agnosticism as transparency of the provided application interface. Besides the adherence to the RDF data format the system also has to provide a data access and update interface following the SPARQL 1.1 [41] standard.

## 4. State of the Art

In the following we look into existing approaches for partially solving the targeted problem. First we consider abstract models to express changes and evolution, like methodologies and vocabularies to manage the decentralized evolution of RDF datasets in section 4.1. Second, we examine implementations dealing with versioning of RDF data in section 4.2. Followed by applications built on top of RDF versioning systems in section 4.3.

### *4.1. Theoretical Foundations and Vocabularies*

Currently, various vocabularies exist that allow the description of provenance. As a *World Wide Web Consortium* (W3C) Recommendation, the PROV ontology (PROV-O) [34] is the de-facto standard for the representation and exchange of domain-independent provenance. The *Open Provenance Model* [37] predates the PROV-O, but both use very similar approaches as their core components. Both vocabularies enable the description of provenance data as relations between *agents*, *entities*, and *activities* or their respective equivalent.

Another popular standard for general-purpose metadata is *Dublin Core* respective the *Dublin Core Metadata Terms* [14]. The main difference to the prior ontologies is in their perspective on expressing provenance. Both vocabularies provide means to express provenance metadata. While the PROV-O is more focused on activities that lead to a specific entity, Dublin Core focuses on the resulting entities.

One advantage of using domain-independent vocabularies as a core is their applicability to systems and tools that operate without any domain-specific knowledge. PROV-O-Viz[16] is an example of a visualization tool only working with the data expressed according to the PROV ontology.

[16] http://provoviz.org/

Berners-Lee and Connolly [9] give a general overview on the problem of synchronization and how to calculate delta on RDF graphs. This work considers the transfer of changes to datasets by applying patches. They introduce a conceptual ontology that describes patches in *a way to uniquely identify what is changing* and *to distinguish between the pieces added and those subtracted*.

Haase and Stojanovic [26] introduce their concept of ontology evolution as follows:

> Ontology evolution can be defined as the timely adaptation of an ontology to the arisen changes and the consistent management of these changes. [...] An important aspect in the evolution process is to guarantee the consistency of the ontology when changes occur, considering the semantics of the ontology change.

This paper has its focus on the linear evolution process of an individual dataset, rather than the decentralized evolution process. To deal with inconsistency resp. consistency they define three levels: *structural*, *logical*, and *user-defined* consistency. In the remainder of the paper they mainly focus on the implications of the evolution with respect to OWL (DL) rather than a generic approach.

Auer and Herre [7] propose a generic framework to support the versioning and evolution of RDF graphs. The main concept introduced in this work is the concept of *atomic graphs*, which provides a practical approach to deal with blank nodes in change sets. Additionally they introduce a formal hierarchical system to structure a set of changes and evolution patterns that lead to the changes of a knowledge base.

### *4.2. Practical Knowledge Base Versioning Systems*

Table 1 provides an overview of the related work and compares the approaches presented with regard to different aspects. One aspect to categorize versioning systems is its archiving policy. Fernández et al. [17] define three archiving policies. IC - *Independent Copies*, where each version is stored and managed as a different, isolated dataset. CB - *Change-based* (delta), where differences between versions are computed and stored based on a *basic language of changes* describing the change operations. TB - *Timestamp-based* where each statement is annotated with is temporal validity. These three archiving policies do not cover the full range of possible archiving systems and thus additionally we need to define the archiving policy FB - *Fragment-based*. This policy stores snapshots of each changed fragment of an archive. Depending on the requirements fragments can be defined at any level of granularity (e.g. resources, subgraphs, or individual graphs in a dataset). An index is maintained that references the fragments belonging to a version of the dataset. This approach addressed the issue of IC of fully repeating all triples across versions. In contrast to CB it is not necessary to reconstruct individual versions by applying the change operations.

Besides the used archiving policy we compare the systems if they allow collaboration on datasets with multiple graphs (*quad support*, REQ 7) and if it is possible to access individual versions on the repository (*random access*, REQ 5). Finally, we want to find out how the existing systems support the solution of our three basic requirements *Support Divergence* (REQ 1), *Conflate Diverged States* (REQ 2), and *Synchronize* (REQ 3) by allowing to create multiple branches, merge branches, and create distributed setups for collaboration using push and pull mechanisms.

TailR as presented by Meinhardt et al. [35] is a system to preserve the history of arbitrary RDF datasets on the web. It follows a combined delta and snapshot storage approach. The system is comparable to the approach presented by Frommhold et al. [21], as both systems are linear change tracking systems. None of the systems provides support for branches to allow independent evolution of RDF graphs.

Another approach is implemented by *stardog*[17], a triple store with integrated version control capabilities. The versioning module provides functionality to tag and calculate the difference between revisions[18]. Snapshots contain all named graphs from the time the snapshot was taken. RDF data and snapshots are stored in a relational database. The current state of the database can be queried via a SPARQL interface. While older states of the database can be restored, to our knowledge, they cannot be queried directly. An additional graph containing the version history is provided.

Cassidy and Ballantine [12] present a version control system for RDF graphs based on the model of *Darcs*. Their approach covers the versioning operations *commute*, *revert*, and *merge*. Even though *Darcs* is considered a DVCS in contrast to other DVCS the presented approach only supports linear version tracking. Further the merge operation is implemented using patch commutation this requires history rewriting and thus loses the context of the original changes.

Graube et al. [24] propose the R43ples approach that uses named graphs to store revisions as deltas. To express the provenance information it uses the RMO vocabulary[19], which is an extended and more domain-specific version of the PROV ontology. To query and update the triple store an extended SPARQL protocol language is introduced.

R&Wbase by Vander Sande et al. [45] is a tool for versioning an RDF graph. It tracks changes that are stored in individual named graphs that are combined on query time, this situation makes it impossible to use the system to manage RDF datasets with multiple named graphs. The system also implements an understanding of coexist-ing branches within a versioning graph, which is very close to the concept of Git.

[17] `http://stardog.com/`

[18] `http://www.stardog.com/docs/#_versioning`, `https://github.com/stardog-union/stardog-examples/blob/d7ac8b5/examples/cli/versioning/README.md`

[19] `https://github.com/plt-tud/r43ples/blob/master/doc/ontology/RMO.ttl`

| Approach | Archiving Policy | Quad Support | Random Access | Branches/Merge | Synchronize (Push/Pull) |
|---|---|---|---|---|---|
| Frommhold et al. [21] | CB | yes | no | no[d] | no |
| Meinhardt et al. [35] | hybrid (IC and CB) | no[a] | yes | no[d] | (yes)[h] |
| *stardog* | IC | yes | no | no | no |
| Cassidy and Ballantine [12] | CB | no | no | no/(yes)[d,e] | (yes)[i] |
| Vander Sande et al. [45] | TB | no[b] | yes | yes/(yes)[f] | no |
| Graube et al. [24] | CB | yes[b,c] | yes | yes/(yes)[g] | no |
| *dat* | FB (*chunks*) | n/a | yes | no | yes |

[a] The granularity of versioning are repositories; [b] The context is used to encode revisions; [c] Graphs are separately put under version control; [d] Only linear change tracking is supported; [e] If a workspace of duplicated subsequent patches can be applied to the original copy this is called merge in this system; [f] Naive merge implementation; [g] The publication mentions a merging interface; [h] No pull mechanism but history replication via memento API; [i] Synchronizations happens by exchanging patches

Table 1: Comparison of the different (D)VCS systems for RDF data. Custom implementations exist for all of these systems and they are not re-using existing VCSs. At the level of abstraction all of these systems can be located on the *data interchange* layer.

In the *dat*[20] project a tool to distribute and synchronize data is developed, the aim is to synchronize any file type peer to peer. It has no support to manage branches and merge diverged versions and is not focusing on RDF data.

Looking at the above knowledge base versioning systems as listed in table 1, it is clear that only two of them can fulfill the requirement *Support Divergence* by providing a branching model, namely R&Wbase [45] and R43ples [24]. Even so they currently only have very limited support for merge operations to fulfill the requirement *Conflate Divergence*. Moving on to the support for the requirement *Synchronize* we see TailR [35], the approach by Cassidy and Ballantine [12] and *dat*. Given the limited support for RDF, we can ignore *dat* while TailR and the approach by Cassidy and Ballantine do not bring support for a branching system and thus cannot fulfill the first two requirements. One can argue that it is more likely to extend a system with proper support for branching and merging with an appropriate synchronization system than the other way around. Because once all conflicts are resolved and the conflation is performed locally only the storage structure needs to be transferred. Thus the remaining relevant related work is R&Wbase [45] and R43ples [24].

### *4.3. Applications to Exploit Knowledge Versioning*

The Git4Voc, as proposed by Halilaj et al. [27] is a methodology and collection of best practices for collaboratively creating RDF vocabularies using Git repositories. To support vocabulary authors in the process of creating RDF and OWL vocabularies Git4Voc is implemented using pre- and post-commit hooks to validate the vocabulary and generate documentation. To validate the vocabulary specification, a combination of local and online tools is used. In preparation of the presented Git4Voc system, Halilaj et al. have formulated important requirements for collaboration on RDF data. We have partially incorporated these requirements in section 3. Based on Git4Voc Halilaj et al. have created the VoCol [27] system as an integrated development environment for vocabularies. For VoCol, the three core activities *modeling*, *population*, and *testing* are formulated. VoCol, as well as Git4Voc are not focused on providing a versioning system for RDF data in general, but rather tools built on top of a versioning system to specifically support the development of vocabularies.

The Git2PROV[21] tool [15] allows to generate a provenance document using the PROV ontology for any public Git repository. It can be used as a web service or can be executed locally. Since our aim is to provide provenance for RDF data on graph- and triple-level Git2PROV is not suited as a component since it is only able to handle provenance on a per-file-level.

## 5. Preliminaries

In the following we give a brief overview and introduction to technologies and design considerations that are relevant for our methodology. At first we introduce the DVCS *Git* by describing its general architecture and some technological details relevant for the realization of our approach in section 5.1. Then we discuss design considerations regarding the storage and serialization of RDF data in section 5.2 and an approach to support blank nodes in section 5.3.

[20] http://dat-data.com/

[21] http://git2prov.org/

### 5.1. Git

Git[22] is a DVCS designed to be used in software development. It is used to manage over 64 million projects on github[23] and is also used on other platforms, such as bitbucket or gitlab, is hosted on self controlled servers, and is used in a peer to peer manner. Git offers various branching and merging strategies, and synchronizing with multiple remote repositories. Due to this flexibility, best practices and workflows have been developed to support software engineering teams with organizing different versions of a programs source code, such as e.g. *gitflow*[24] and the *Forking Workflow*[25].

In contrast to other VCS such as Subversion or CVS[26], Git is a DVCS. As such, in Git, users work on a local version of a remote Git repository that is a complete clone of the remote repository. Git operations, such as *commit*, *merge*, *revert*, as well as (interactive) *rebase* are executed on the local system. Out of the box, Git already provides the capability to store provenance information alongside a commits. The repository contains commits that represent a certain version of the working directory. Each version of the working directory contains the current state of its files at the given version. Even so Git is mainly intended to work with text files, its storage system does not distinguish between text and arbitrary binary files. Files are stored as binary large objects (*blob*) and are referenced by commits, while equal files are stored only once.

Thus, to extend our provenance information we have to dig a little deeper into Git's internal structure. The internal storage structure of Git is a file-system based key-value store. Git uses different types of objects to store both structural information and data in files. The types used by Git to organize its data are *blobs*, *trees*, and *commits*, the types are linked as shown in fig. 2. The individual data items are addressed with their sha1-hash[27]. The sha1-hash is also used as commit ID resp. blob ID.

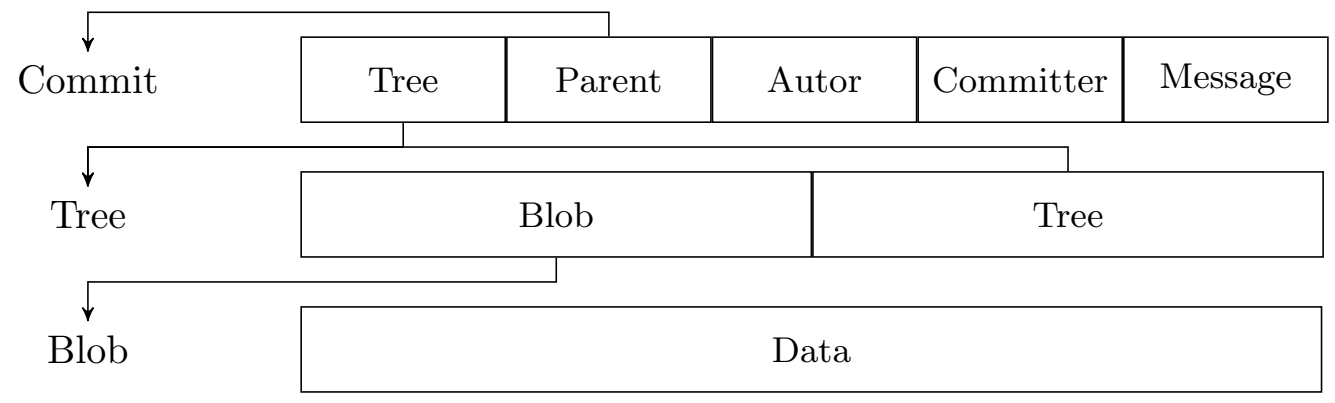

Figure 2: Internal structure used by Git.

The content of any file that is put under version control is stored as *blob* while a folder are stored as *tree*. A *tree* is a list of references to other *trees* and *blobs*. Each revision is represented by a *commit*-object consisting of metadata, references to parent commits and a reference to a *tree*-object that is considered as root. References such as *branches* and *tags* are simply files within Git, pointing to a commit as their entry point to the revision history. With the Git transfer protocol it is possible to synchronize distributed repositories without the need for a central instance.

### 5.2. Serialization of RDF Data

RDF 1.1 specifies multiple different formats that can be used to serialize RDF graphs (RDF/XML[28], Turtle[29], RDFa[30], N-Triples[31]) and RDF datasets (TriG[32], JSON-LD[33], N-Quads[34]). RDF graphs and RDF datasets can be serialized in different formats and thus the same RDF statements can result in completely different textual representations and the resulting file size can vary. Even the same graph or dataset serialized twice in the same serialization format can be textually different. To allow a better readability and processability of the differences between two versions in the VCS (cf. section 3 "Deltas Among Versions"), we have to find an easy to compare default serialization format. For our approach we have decided to use the N-Quads serialization [11] in Git repositories. N-Quads is a line-based, plain text format that represents one statement per line. Since Git is also treating lines as atoms it will automatically treat statements in N-Quads as atomic units In contrast to N-Triples, N-Quads supports the encoding of complete RDF datasets (multiple graphs). N-Triples is a subset of N-Quads, by only using the default graph. Another candidate would be TriG (Turtle extended by support for RDF datasets), in contrast to N-Quads one line does not necessarily represent one statement. Also due to the abbreviation features (using `;` or `,` as delimiter) as well as multi line literals, automatic line merges can destroy the syntax. Similar problems would occur with the other serialization formats listed above. To further ensure stability and comparability of the files we maintain a sorted list of statements during the serialization.

Halilaj et al. [27] propose the usage of Turtle in Git repositories, to address the requirement to be editor agnostic. Since a transformation to any other serialization format is possible, e.g. using rapper[35] or Jena RIOT[36], our approach does not put additional constraints on the usage of the serialization format in an editor application. Further as stated above, we find N-Quads to be of better fit for Git versioning than Turtle.

[22] https://git-scm.com/
[23] https://github.com/about, 2017-08-15
[24] http://nvie.com/posts/a-successful-git-branching-model/
[25] https://www.atlassian.com/git/tutorials/comparing-workflows/forking-workflow
[26] Concurrent Versions System, http://savannah.nongnu.org/projects/cvs
[27] Secure Hash Algorithm
[28] https://www.w3.org/TR/2014/REC-rdf-syntax-grammar-20140225/
[29] https://www.w3.org/TR/2014/REC-turtle-20140225/
[30] https://www.w3.org/TR/2015/NOTE-rdfa-primer-20150317/
[31] https://www.w3.org/TR/2014/REC-n-triples-20140225/
[32] https://www.w3.org/TR/2014/REC-trig-20140225/
[33] https://www.w3.org/TR/2014/REC-json-ld-20140116/
[34] https://www.w3.org/TR/2014/REC-n-quads-20140225/
[35] http://librdf.org/raptor/rapper.html
[36] https://jena.apache.org/documentation/io/

*5.3. Blank Nodes in Versioning*

Using RDF as an exchange format, blank nodes are still a problem we have to deal with. Blank nodes are identifiers with a local scope and so might be different for each participating platform. The recommendation of RDF 1.1 suggests replacing blank nodes with IRIs [13], which is called skolemization. Replacing all inserted blank nodes with skolem-IRIs would alter the stored dataset. Our preferred solution is thus to break down all operations on the dataset to atomic graphs, as proposed by Auer and Herre [7].

## 6. Definitions

In this section we introduce a formalization to express changes to RDF graphs based on additions and deletions of atomic subgraphs. This foundational formal model is used to describe the more complex operations in sections 7 and 8. As it is commonly used in RDF and as it is also defined in [13] we define an *RDF graph* resp. just *graph* as a set of RDF triples. With the exception, that we consider isomorphic sub-graphs as identical and de-duplicate these sub-graphs during our operations.[37] An *RDF dataset* resp. just *dataset* is a collection of RDF graphs as defined in [13].

According to Auer and Herre [7] an *Atomic Graph* is defined as follows:

**Definition 1 (Atomic Graph).** A graph is called atomic if it can not be split into two nonempty graphs whose respective sets of blank nodes are disjoint.

This implies that all graphs containing exactly one statement are atomic. Furthermore a graph is atomic if it contains a statement with at least on blank node and each pair of occurring blank nodes is connected by a sequence of statements where subject and object are blank nodes. If one of these statements additionally contains a second blank node, the same takes effect for this blank node recursively. A recursive definition of *Atomic Graphs* is given under the term *Minimum Self-Contained Graph* (MSG) by Tummarello et al. [44].

Let $\mathbb{A}$ be the set of all *Atomic Graphs* and let $\approx$ be the equivalence relation such that $G \approx H$ holds for any $G, H \in \mathbb{A}$ iff $G$ and $H$ are isomorphic as defined for RDF graphs in [13]. Essentially two graphs are isomorphic in this sense if a bijection between these graphs exists that is the identity mapping for non-blank nodes and predicates and a bijection between blank-nodes. By $\mathcal{P} := \mathbb{A}/\approx$ we denote the quotient set of $\mathbb{A}$ by $\approx$. We assume a canonical labeling for blank nodes for any graph. The existence of such a labeling has been shown by Hogan [30]. Thus a system of representatives of $\mathcal{P}$ is given by the set $\mathbb{P} \subset \mathbb{A}$ of all canonically labeled atomic graphs.

Based on this we now define the *Canonical Atomic Partition* of a graph as follows:

**Definition 2 (Canonical Atomic Partition).** Given an RDF graph $G$, let $P_G \subset \mathbb{A}$ denote the partition of $G$ into atomic graphs. We define a mapping $r : P_G \to \mathbb{P}$, such that $r(a) = p$, where $p$ is the canonically labeled representative of $a$.

The *Canonical Atomic Partition* of the graph $G$ is defined as

$$\mathcal{P}(G) := \{r(x) | x \in P_G\}$$

$\mathcal{P}(G) \subset \mathbb{P}$ and especially $\mathcal{P}(G) \subset \mathbb{A}$.

Each of the contained sets consists of exactly one statement for all statements without blank nodes. For statements with blank nodes, it consists of the whole subgraph connected to a blank node and all neighboring blank nodes. This especially means that all sets in the *Atomic Partition* are disjoint regarding the contained blank nodes (cf. [7, 44]). Further they are disjoint regarding the contained triples (because it is a partition).

Since $\mathcal{P}(G)$ is a set of atomic graphs, the union of its elements is a graph again and it is isomorphic to $G$,

$$\bigcup \mathcal{P}(G) \approx G.$$

Because we build a system for distributed collaboration on datasets, we need to find a way to express the changes that lead from one dataset to another. To express these changes we start by comparing two graphs by calculating the *difference*.

**Definition 3 (Difference).** Let $G$ and $G'$ be two graphs, and $\mathcal{P}(G)$ resp. $\mathcal{P}(G')$ the *Canonical Atomic Partitions*.

$$\begin{aligned} C^+ &:= \bigcup (\mathcal{P}(G') \setminus \mathcal{P}(G)) \\ C^- &:= \bigcup (\mathcal{P}(G) \setminus \mathcal{P}(G')) \\ \Delta(G, G') &:= (C^+, C^-) \end{aligned}$$

Looking at the resulting tuple $(C^+, C^-)$ we can also say that the inverse of $\Delta(G, G')$ is $\Delta^{-1}(G, G') = \Delta(G', G)$ by swapping the positive and negative sets.

We now have the tuple of additions and deletions that describes the difference between two graphs. Thus we can say that applying the changes in this tuple to the initial graph $G$, leads to $G'$. Furthermore we can define a *Changeset* that can be applied on a graph $G$ as follows:

**Definition 4 (Changeset).** Given an RDF graph $G$, a changeset is a tuple of two graphs $(C_G^+, C_G^-)$ in relation to $G$, with

$$\begin{aligned} \mathcal{P}(C_G^+) &\cap \mathcal{P}(G) = \emptyset \\ \mathcal{P}(C_G^-) &\subset \mathcal{P}(G) \\ \mathcal{P}(C_G^+) &\cap \mathcal{P}(C_G^-) = \emptyset \\ \mathcal{P}(C_G^+) &\cup \mathcal{P}(C_G^-) \neq \emptyset \end{aligned}$$

[37] This is not the same definition as for lean-graphs [29], but our graphs are similar to lean-graphs regarding the aspect that we eliminate internal redundancy.

Since blank nodes cannot be identified across graphs, there cannot be any additions of properties to a blank node, nor can properties be removed from a blank node. If a change to a statement involving a blank node takes place, this operation is transformed into the removal of one atomic graph and the addition of another atomic graph. Thus $\mathcal{P}(C_G^+)$ and $\mathcal{P}(G)$ have to be disjoint. This means an addition cannot introduce just new statements to an existing blank node. Parallel to the addition a blank node can only be removed if it is completely removed with all its statements. This is ensured by $\mathcal{P}(C_G^-)$ being a subset of $\mathcal{P}(G)$. Simple statements without blank nodes can be simply added and removed. Further since $\mathcal{P}(C_G^+)$ and $\mathcal{P}(C_G^-)$ are disjoint we avoid the removal of atomic graphs that are added in the same changeset and vice versa. Since at least one of $\mathcal{P}(C_G^+)$ or $\mathcal{P}(C_G^-)$ cannot be empty we avoid changes with no effect.

**Definition 5 (Application of a Change).** Given an RDF graph $G$, let $C_G = (C_G^+, C_G^-)$ be a changeset on $G$. The function *Apl* is defined for the arguments $G, C_G$ respective $G, (C_G^+, C_G^-)$ and is determined by

$$Apl\left(G, (C_G^+, C_G^-)\right) := \bigcup \left((\mathcal{P}(G) \setminus \mathcal{P}(C_G^-)) \cup \mathcal{P}(C_G^+)\right)$$

We say that $C_G$ is applied to $G$ with the result $G'$.

## 7. Operations

Based on the formal model to express changes on RDF graphs, we can now introduce our formal syntax to represent versioning operations. Because we aim for version tracking in a decentralized evolution environment linear versioning operation are not enough, moreover our operations have to bring in support for non-linear versioning. The presented operations are *commit* to record changes in section 7.1, *branch* to support divergence in section 7.2, *merge* in section 7.3, and *revert* to undo changes in section 7.4. Each version is related to the respective RDF dataset and thus the versioning operations are related to the respective transformations of the RDF dataset.

### *7.1. Commit*

Figure 3 depicts an initial commit $\mathcal{A}$ without any ancestor resp. parent commit and a commit $\mathcal{B}$ referring to its parent $\mathcal{A}$.

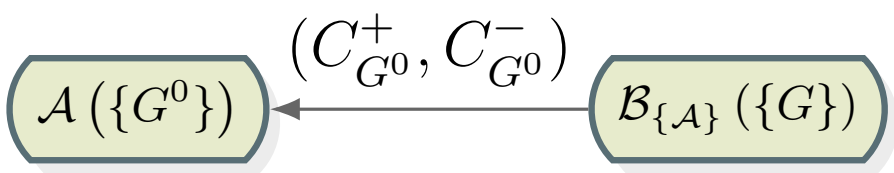

Figure 3: Two commits with an ancestor reference.

Let $G^0$ be a graph under version control. $\mathcal{A}(\{G^0\})$ is a commit containing the graph $G^0$. $G$ will be the new version of $G^0$ after a change $(C_{G^0}^+, C_{G^0}^-)$ was applied on $G^0$; $Apl(G^0, (C_{G^0}^+, C_{G^0}^-)) = G$. Now we create a new commit that contains $G$ and refers to its ancestor commit, from which it was derived: $\mathcal{B}_{\{\mathcal{A}\}}(\{G\})$. Another change applied on $G$ would result in $G'$ and thus a new commit $\mathcal{C}_{\{\mathcal{B}_{\{\mathcal{A}\}}\}}(\{G'\})$ is created. In the further writing, the indices and arguments of commits are sometimes omitted for better readability, while clarity should still be maintained by using distinct letters. Further also the changeset on top of the arrow is omitted if it is obvious.

The evolution of a *graph* is the process of subsequently applying changes to the graph using the *Apl* function as defined in definition 5. Each commit expresses the complete evolution process of a set of graphs, since it refers to its ancestor, which in turn refers to its ancestor as well. Initial commits holding the initial version of the graph are not referring to any ancestor.

### *7.2. Branch*

Since a commit refers to its ancestor and not vice versa, nothing hinders us to create another commit $\mathcal{D}_{\{\mathcal{B}_{\{\mathcal{A}\}}\}}(\{G''\})$. Taking the commits $\mathcal{A}$, $\mathcal{B}_{\{\mathcal{A}\}}$, $\mathcal{C}_{\{\mathcal{B}\}}$, and $\mathcal{D}_{\{\mathcal{B}\}}$ results in a directed rooted in-tree, as depicted in fig. 4. The commit $\mathcal{D}$ is now a new *branch* or *fork* based on $\mathcal{B}$, which is diverged from $\mathcal{C}$. We know that $G \not\approx G'$ and $G \not\approx G''$, while we do not know about the relation between $G'$ and $G''$.

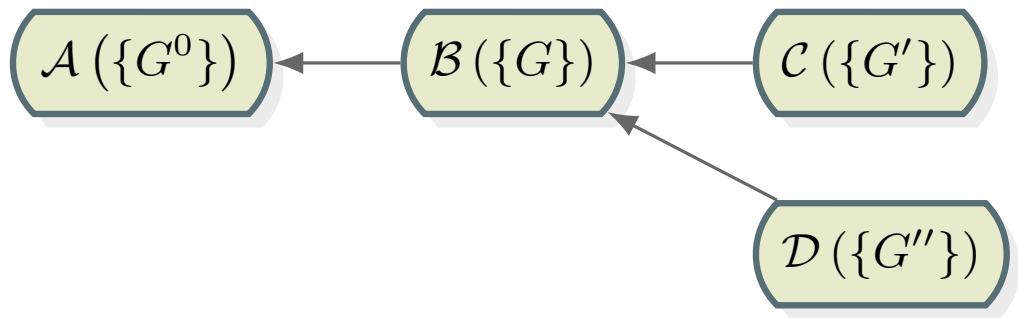

Figure 4: Two branches evolved from a common commit.

Because we do not know anything about the relation between $G'$ and $G''$ we can consider them as independent. From now on the graph $G$ is *independently* evolving in two branches, while *independent* means possibly independent, this means two contributors performing a change do not have to know of each other or do not need a direct communication channel. The contributors could actually communicate, but communication is not required for those actions. Thus by branching a dataset's evolution the contributors can be working from distributed places and no central instance for synchronization is required. We thus define the *branch* as follows:

**Definition 6 (Branching).** Branching is the (independent) evolution of a graph $G$ with two graphs $G_1$ and $G_2$ as result, where $Apl(G, C_1) = G_1$ and $Apl(G, C_2) = G_2$. The changes $C_1$ and $C_2$ might be unequal, but can be the same. The same applies for $G_1$ and $G_2$, they can be different after the independent evolution, but can be similar as well.

### *7.3. Merge Different Branches*

After creating a second branch, the tree of commits is diverged, as shown in the example of fig. 4. We now want to merge the branches, in order to get a version of the graph, containing changes made in those different branches or at least take all of these changes into account. The notation of the merge is defined as follows:

**Definition 7 (Merge of two Evolved Graphs).** Given are two commits $\mathcal{C}_{\{\beta\}}(\{G'\})$ and $\mathcal{D}_{\{\gamma\}}(\{G''\})$. Merging the two *graphs* $G'$ and $G''$ with respect to the change history expressed by the commits $\mathcal{C}$ and $\mathcal{D}$ is a function

$$Merge(\mathcal{C}(\{G'\}), \mathcal{D}(\{G''\})) = \mathcal{M}_{\{\mathcal{C},\mathcal{D}\}}(\{G^{\mu}\})$$

The $Merge$ function takes two commits as arguments and creates a new commit dependent on the input commits, this new commit is called *merge commit*. The graph $G^{\mu}$ is the *merged graph* resulting from $G'$ and $G''$. The merge commit resulting from the merge operation has two ancestor commits that it refers to. If we take our running example the merge commit is $\mathcal{M}_{\{\mathcal{C}_{\{\mathcal{B}\}},\mathcal{D}_{\{\mathcal{B}\}}\}}(\{G^{m}\})$. Taking the commits $\mathcal{A}$, $\mathcal{B}_{\{\mathcal{A}\}}$, $\mathcal{C}_{\{\mathcal{B}\}}$, $\mathcal{D}_{\{\mathcal{B}\}}$, and $\mathcal{M}_{\{\mathcal{C},\mathcal{D}\}}$, we get an acyclic directed graph, as it is depicted in fig. 5.

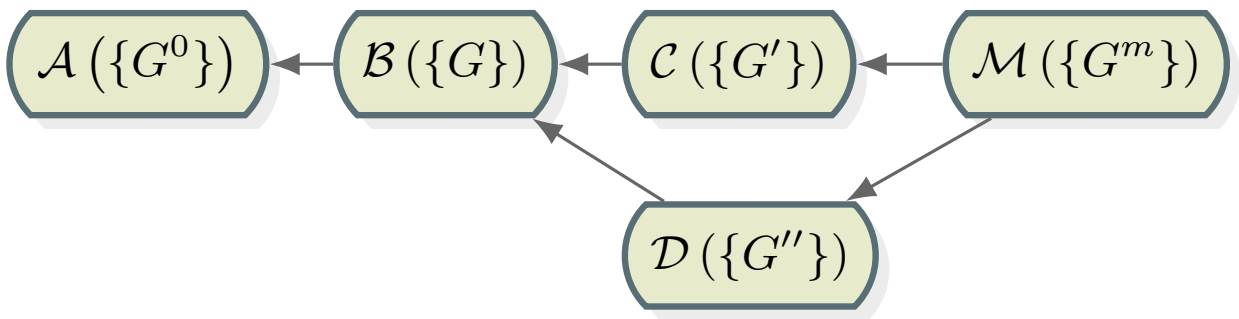

Figure 5: Merging commits from two branches into a common version of the graph.

Note that the definition does not make any assumptions about the ancestors of the two input commits. It depends on the actual implementation of the $Merge$ function whether it is required that both commits have any common ancestors. Furthermore different merge strategies can produce different results, thus it is possible to have multiple merge commits with different resulting graphs but with the same ancestors. Possible merge strategies are presented in section 8.

### 7.4. Revert a Commit

Reverting the commit $\mathcal{B}_{\{\mathcal{A}\}}(\{G\})$ is done by creating an inverse commit $\mathcal{B}^{-1}_{\{\mathcal{B}\}}(\{\tilde{G}^0\})$ (while the commit $\mathcal{A}$ is specified as $\mathcal{A}(\{G^0\})$). This inverse commit is then directly applied to $\mathcal{B}$. The resulting graph $\tilde{G}^0$ is calculated by taking the inverse difference $\Delta^{-1}(G^0, G) = \Delta(G, G^0)$ and applying the resulting change to $G$. After this operation $\tilde{G}^0 = G^0$.

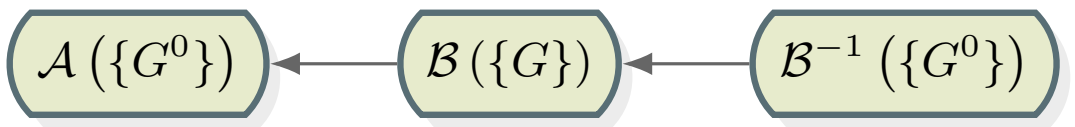

Figure 6: A commit reverting the previous commit.

A versioning log containing three commits is shown in fig. 6. The last commit reverts its parent and thus the graph in $\mathcal{B}^{-1}$ is again equal or at least equivalent to the graph in the first commit ($\mathcal{A}$). While it is obvious how to revert the previous commits, it might be a problem if other commits exist between the commit to be reverted and the current top of the versioning log. In this case a merge is applied (cf. section 8). For this merge, the merge *base* is the commit to be reverted, branch $\mathcal{A}$ is the parent commit of the commit that is to be reverted, and branch $\mathcal{B}$ the current latest commit. Arising conflicts when reverting a commit can be resolved in the same way, as for merge commits.

## 8. Merge Strategies

Since we are not only interested in the abstract branching and merging model of the commits, we want to know what a merge operation means for the created graph in the commit. In the following we present some possible implementations of the merge operations. Note, that *merging* in this context is not—generally—to be understood as in the *RDF 1.1 Semantics Recommendation* [29] as the union of two graphs.

### 8.1. Union Merge

Merging two *graphs* $G'$ and $G''$ could be considered trivially as the union operation for the two graphs: $G' \cup G'' = G'''$. This merge – as mentioned above – is well defined in the *RDF 1.1 Semantics Recommendation* [29] in section “4.1 Shared blank nodes, unions and merges”. But this operation would not take into account the actual change operations leading to the versions of the graphs. Furthermore the union merge does not allow the implementation of conflict detection or resolution operations. The union merge might be intended in situations, where data is only added from different locations.

### 8.2. All Ours/All Theirs

Two other merge strategies that would not produce merge conflicts are *ours* and *theirs*. These merge strategies just take the whole graph $G' =: G'''$ or $G'' =: G'''$ as new version, while they are ignoring the other graph respectively. This strategy might be chosen to completely discard changes from a certain branch.

### 8.3. Three-Way-Merge: An Unsupervised Approach to Merge Branched Knowledge Bases

A methodology used in DVCS for software source code files, such as Git and Mercurial is the *three-way-merge*[38]. The merge consists of three phases, (1) finding a common *merge base* for the two commits to merge, (2) comparing the files between the *merge base* and the individual branches and inferring which lines where added and removed, and (3) creating a merged version by *combining* the changes made in the two branches.

[38]How does Git merge work: `https://www.quora.com/How-does-Git-merge-work`, 2016-05-10

| A $G'$ | B $G''$ | base $G$ | result $G^m$ | $\Delta(G,G')$ $C_A^+$ | $C_A^-$ | $\Delta(G,G'')$ $C_B^+$ | $C_B^-$ | |
|---|---|---|---|---|---|---|---|---|
| | | | | | | | | Non existing statements |
| X | X | X | X | | | | | Atomic graph existent in all graphs will also be in the result |
| X | | | X | X | | | | An atomic graph added to $G'$ is also added to the result |
| | X | | X | | | X | | An atomic graph added to $G''$ is also added to the result |
| | X | X | | | X | | | An atomic graph removed from $G'$ is also not added to the result |
| X | | X | | | | | X | An atomic graph removed from $G''$ is also not added to the result |
| X | X | | X | X | | X | | An atomic graph added to both branches is also added to the result |
| | | X | | | X | | X | An atomic graph removed from both branches is also not added to the result |

Table 2: Decision table for the different situations on a three-way-merge (X = atomic graph exists, *empty* = atomic graph does not exist).

Implementing this function for *combining* the changes is the actual problem and task of the selected merge algorithm. A merge algorithm can take several aspects into account when deciding, whether to include a line into the merged version or not. If the algorithm cannot decide on a certain line, it produces a merge conflict. For source code files in Git this is for instance the case, if two close-by lines where changed in different branches. Since the order of source code lines is crucial, a merge conflict is produced.

We transform this situation to RDF datasets. We take into account the versions of the graphs in the two commits to be merged $\mathcal{C}_{\{\mathcal{B}\}}(\{G'\})$, $\mathcal{D}_{\{\mathcal{B}\}}(\{G''\})$. To find the merge base (1) this strategy relies on the existence of a common ancestor $\delta$, such that for $\mathcal{C}$ and $\mathcal{D}$ there must exist an ancestor path $\mathcal{C}_{\{\cdot\cdot_{\{\delta,\dots\}}\}}$ resp. $\mathcal{D}_{\{\cdot\cdot_{\{\delta,\dots\}}\}}$ to $\delta$. In our case we find the most recent common ancestor $\delta := \mathcal{B}(\{G\})$. Now we have to compare (2) the graphs between the *merge base* and the individual branches and calculate their difference:

$$(C^+_{\mathcal{C}}, C^-_{\mathcal{C}}) = \Delta(G, G')$$
$$(C^+_{\mathcal{B}}, C^-_{\mathcal{B}}) = \Delta(G, G'')$$

In contrast to source code files, there is no order relevant in the RDF data model. Thus we can just take the resulting sets of the comparison and merge them into a new version (3) as follows:

$$G''' = \bigcup \left( (\mathcal{P}(G') \cap \mathcal{P}(G'')) \cup \mathcal{P}(C^+_{\mathcal{C}}) \cup \mathcal{P}(C^+_{\mathcal{D}}) \right)$$

A more visual representation of the three-way-merge is given as decision matrix in table 2. The table shows in the first three columns, all combinations of whether a statement is included in one of the two branches and their *merge base*. The fourth column shows whether a statement is present in the merge result as defined for the three-way-merge. The other four columns visualize the presence of a statement in the deltas between the two branches and the *merge base* respectively.

This merge strategy purely takes the integrity of the RDF data model into account. This especially means, that semantic contradictions have to be dealt with in other ways. One possibility to highlight possible contradictions as merge conflicts is the context merge strategy (cf. section 8.4). But also beyond a merge operation, semantic contradictions within the resulting RDF graph can be handled using continuous integration tools, as pointed out in section 11.

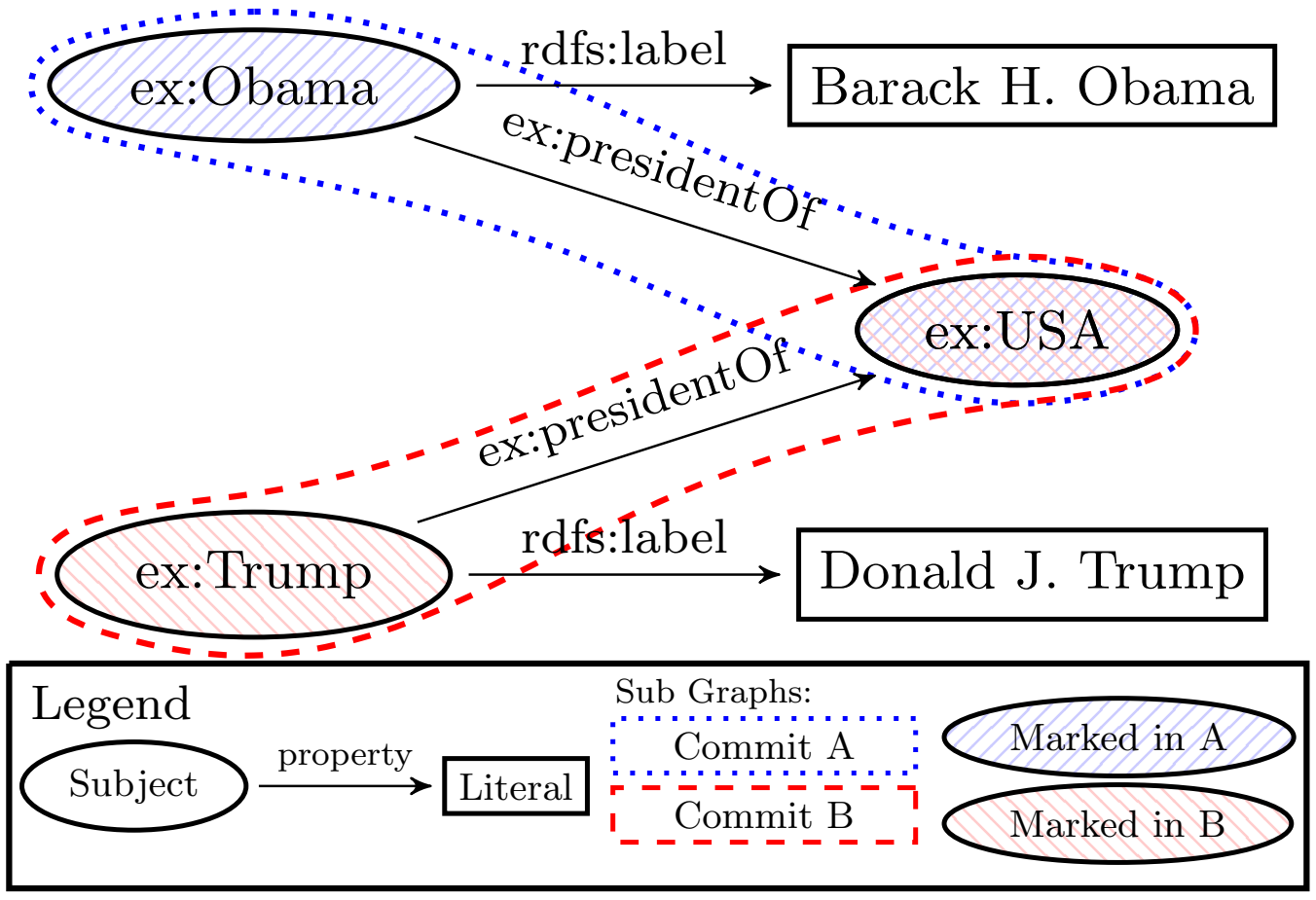

Figure 7: An example for a conflict using the Context Merge

### 8.4. *Context Merge: A Supervised Approach to Identify Conflicts*

Since the *Three-Way-Merge* does not produce any merge conflicts, it can happen, that during a merge semantic conflicts are introduced. Even though the result is a valid RDF graph two statements could contradict each other. Looking at fig. 7 we see, that in commit A the statement *Obama is president of the USA* was introduced, while in commit B the statement *Trump is president of the USA* is added. The result of the *Three-Way-Merge* would be the whole graph as shown in fig. 7.

Since we do not want to apply any specially encoded semantic rules to identify conflicts we have to rely on the pure RDF data model. Thus we can take into account the semantics of nodes and edges and the semantics of additions and deletions as we have seen them in the *Three-Way-Merge*. Let us transfer the principle of producing conflicts in a *Three-Way-Merge* as implemented in Git from source code files to graphs. In files a conflict is produced, as soon, as the merge strategy cannot decide on two lines coming from two different commits, in which order they should be stored. The lines thus are overlapping. The *Context Merge* for RDF is based on the *Three-Way-Merge* in the way, that it performs the merge by taking into account the two commits and their *merge base*. In contrast to the merge as defined in section 8.3 it produces *merge conflicts*, as soon as the changes of both merged commits overlap at a node. If an atomic graph is added resp. removed in both commits there is obviously no contradiction and hence no conflict. The merge process marks each subject and object of an added or removed atomic graph in the graph with the belonging to its originating commit (cf. fig. 7, Sub Graphs). As soon as a node is marked for both commits, this node is added to the list of conflicting nodes. The user is presented with all atomic graphs of the both change sets that contain nodes that are listed as conflicting nodes.

The possible changes in question to be marked as a conflict are those statements[39] that where added or removed in one of the merged branches. Looking at table 2 we see, that both branches agree on the last two lines, while they do not agree on lines 3 to 6.

To perform a context merge we first need to calculate the change sets:

$$\begin{aligned}(C_{\mathcal{C}}^{+}, C_{\mathcal{C}}^{-}) &= \Delta(G, G')\\ (C_{\mathcal{B}}^{+}, C_{\mathcal{B}}^{-}) &= \Delta(G, G'')\end{aligned}$$

As a precondition to identify the conflicts we thus only need the statements, where two branches do not agree. We denote the set of statements present (resp. absent) in $\mathcal{C}$, where $\mathcal{C}$ and $\mathcal{B}$ disagree as follows (disagreed statements):

$$\begin{aligned}\tilde{C}_{\mathcal{C}\setminus\mathcal{B}}^{+} &= C_{\mathcal{C}}^{+} \setminus C_{\mathcal{B}}^{+}\\ \tilde{C}_{\mathcal{C}\setminus\mathcal{B}}^{-} &= C_{\mathcal{C}}^{-} \setminus C_{\mathcal{B}}^{-}\end{aligned}$$

Also to identify the nodes of a statement, the set of all subject nodes and object nodes of $G$ is defined as:

$$N(G) := \{x \mid \exists p, o : (x, p, o) \in G \vee \exists s, p : (s, p, x) \in G\}$$

The set of potentially conflicting nodes is the intersection of the nodes of the disagreed statements:

$$I_N = N\left(\tilde{C}_{\mathcal{C}\setminus\mathcal{B}}^{+} \cup \tilde{C}_{\mathcal{C}\setminus\mathcal{B}}^{-}\right) \cap N\left(\tilde{C}_{\mathcal{B}\setminus\mathcal{C}}^{+} \cup \tilde{C}_{\mathcal{B}\setminus\mathcal{C}}^{-}\right)$$

Now we have to find the respective statements that have to be marked as conflicts, thus the set of all statements in $G$ which contain a node in $I$ is defined on $G$ and $I$ as:

$$\lightning_I(G) := \{(s, p, o) \in G \mid s \in I \vee o \in I\}$$

Thus we have the following sets of potentially conflicting statements:

$$\lightning_{I_N}\left(\tilde{C}_{\mathcal{C}\setminus\mathcal{B}}^{+}\right), \lightning_{I_N}\left(\tilde{C}_{\mathcal{C}\setminus\mathcal{B}}^{-}\right), \lightning_{I_N}\left(\tilde{C}_{\mathcal{B}\setminus\mathcal{C}}^{+}\right), \lightning_{I_N}\left(\tilde{C}_{\mathcal{B}\setminus\mathcal{C}}^{-}\right)$$

While the set of statements that will be contained in the result without question is:

$$\begin{aligned}&(\mathcal{P}(G') \cap \mathcal{P}(G''))\\ &\cup \left(C_A^{+} \setminus \lightning_{I_N}\left(\tilde{C}_{\mathcal{C}\setminus\mathcal{B}}^{+}\right)\right) \cup \left(C_B^{+} \setminus \lightning_{I_N}\left(\tilde{C}_{\mathcal{B}\setminus\mathcal{C}}^{+}\right)\right)\end{aligned}$$

Assuming a function $R$ that gives us the conflict resolution after a user interaction, we end up with a merge method as follows:

$$\begin{aligned}G''' = \bigcup\Bigg(&(\mathcal{P}(G') \cap \mathcal{P}(G''))\\ &\cup \left(C_A^{+} \setminus \lightning_{I_N}\left(\tilde{C}_{\mathcal{C}\setminus\mathcal{B}}^{+}\right)\right) \cup \left(C_B^{+} \setminus \lightning_{I_N}\left(\tilde{C}_{\mathcal{B}\setminus\mathcal{C}}^{+}\right)\right)\\ &\cup R\Bigg(\lightning_{I_N}\left(\tilde{C}_{\mathcal{C}\setminus\mathcal{B}}^{+}\right), \lightning_{I_N}\left(\tilde{C}_{\mathcal{C}\setminus\mathcal{B}}^{-}\right),\\ &\qquad \lightning_{I_N}\left(\tilde{C}_{\mathcal{B}\setminus\mathcal{C}}^{+}\right), \lightning_{I_N}\left(\tilde{C}_{\mathcal{B}\setminus\mathcal{C}}^{-}\right)\Bigg)\Bigg)\end{aligned}$$

This merge strategy relies on the local context in graphs by interpreting subjects and objects of statements as nodes, while predicates are seen as edges. In a context, where a different treatment of predicates is needed, the method can be extended to also mark statements that identify overlapping usage of predicates as well.

## 9. Versioning System

Following the foundational work of the last chapters in this chapter we describe the architecture of our system. An overview with the individual components is given in fig. 8. The Quit API provides query and update interfaces following the Semantic Web standards SPARQL and RDF, as well as interfaces to control the Git repository, which are described in section 9.1. The storage and access facilities for provenance information regarding the aspects *Content*, *Management*, and *Use* (cf. [25]) are described in sections 9.2 and 9.3.

### *9.1. Quit API*

As an interface accessible to other applications, we provide a standard SPARQL 1.1 endpoint. The endpoint supports SPARQL 1.1 *Select* and *Update* to provide a read-/write interface on the versioned RDF dataset. To perform additional Git operations we provide an additional maintenance interface.

The prototypical implementation of the Quit Store[40] is developed using Python[41], with the Flask API[42] and the RDFlib[43] to provide a *SPARQL 1.1 Interface* via HTTP.

[39]For simplicity, we deal with statements in the following definitions and formulas rather then atomic graphs. To transfer the method to atomic graphs, a slightly changed definition of $N(G)$ and $\lightning_I(G)$ with respect to atomic graphs is needed.

[40]`https://github.com/AKSW/QuitStore`

[41]`https://www.python.org/`

[42]`http://flask.pocoo.org/`

[43]`https://rdflib.readthedocs.io/en/stable/`

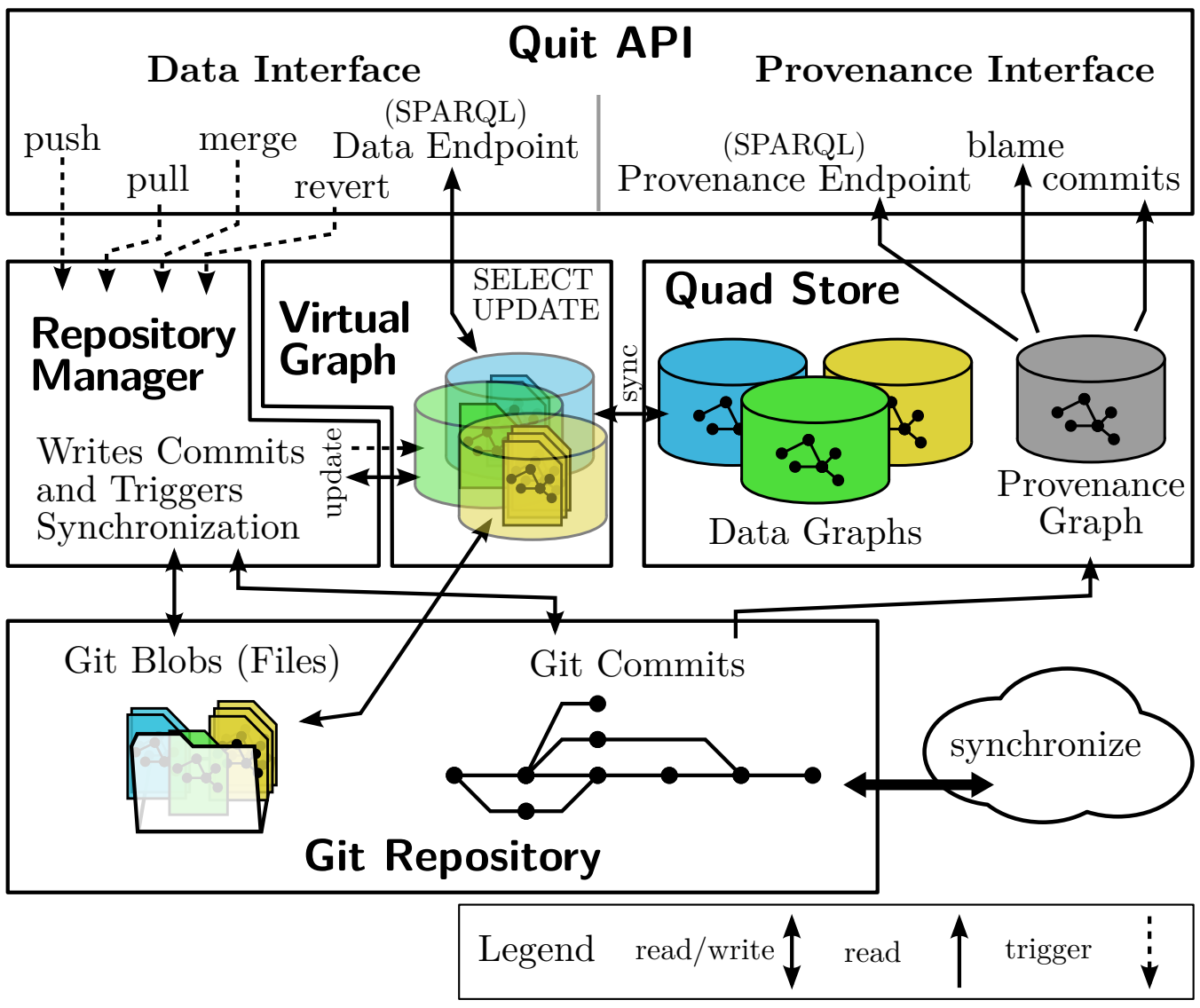

Figure 8: The components of the Quit Store.

The operations on the Git Repository are pursued using the libgit2[44] bindings for Python, pygit2[45]. The underlying storage of the RDF dataset is implemented by an in-memory *Quad-Store* as provided by the RDFlib and a local Git repository that is kept in sync with the corresponding named graphs in the store. Every graph is stored in a canonicalized N-Quads serialization (cf. section 5.2).

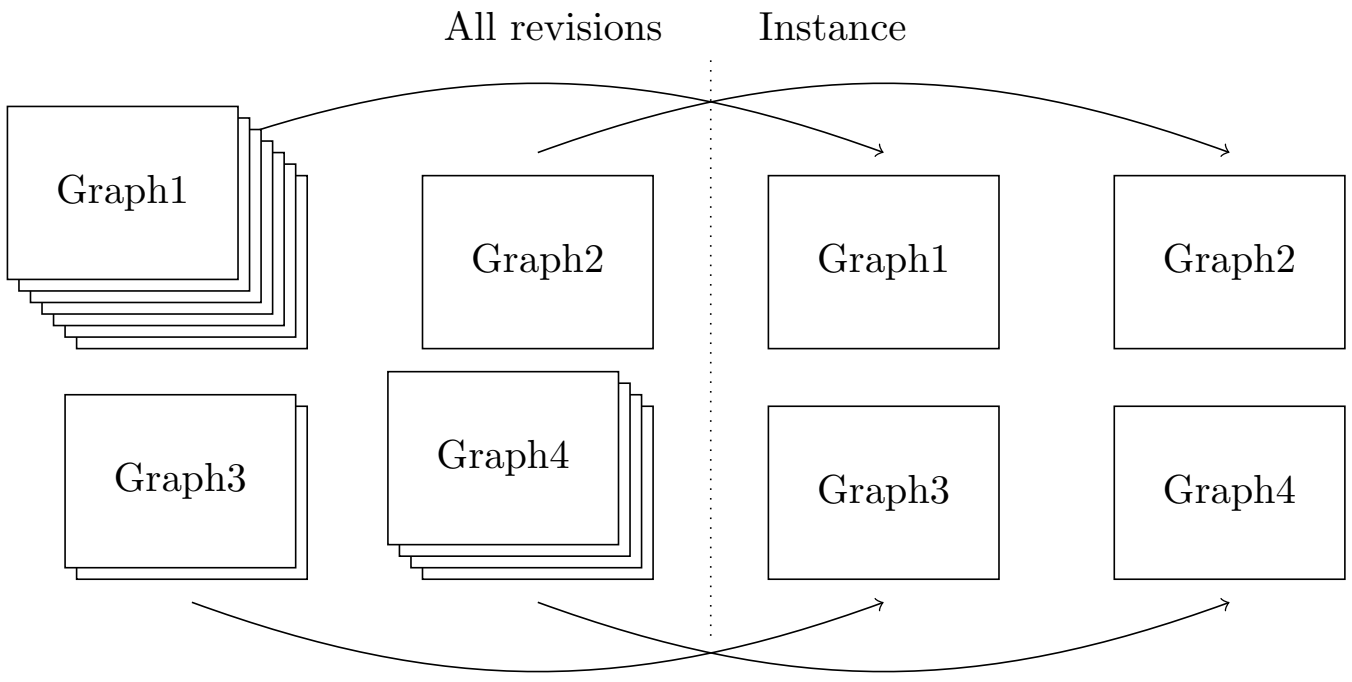

Figure 9: Creating a dataset from all available revisions of named graphs.

*Dataset Access.* To provide access to the versioned RDF dataset multiple virtual endpoints are provided. The default endpoint provides access to the current latest version of the dataset (cf. listing 1, line 1). For each branch (cf. listing 1, line 2) and for each version in the history (cf. listing 1, line 3) of the dataset an additional endpoint is provided. To implement this we build on the basic concept of Git that is to reuse as many of its objects as possible for a new revision. In a Git repository snapshots of modified files, instead of a snapshot of the complete dataset, are stored as (*Git Blobs*). This corresponds to the archiving policies *fragment based* (FB) as defined in section 4.2. Exploiting this storage structure, we can randomly checkout any Git commit in linear time. This allows us to lookup all versions of a graph from Git's internal tree structure to create a virtual dataset (*Virtual Graph*), shown in fig. 9, containing the state of all graphs at that commit and run queries against it. The *Virtual Graph* (cf. fig. 8) thus represents the complete history of the dataset. For better accessibility it is always keeping the latest version of each graph and thus the latest dataset available (cf. fig. 9) in the *Quad Store*. Additionally it maintains the most recently used graphs in the *Quad Store* as well. Received SPARQL-queries are forwarded to the *Virtual Graph* that distinguishes between Update and Select Queries. For Select Queries it ensures, that all graphs of the respective dataset are available in the *Quad Store* and then evaluates the query against it. Update Queries are also evaluated on the internal *Quad Store*. The effective changes on the dataset are then applied to the corresponding *Git Blob* of the ancestor commit. The resulting *Git Blob* is then enclosed in a new Git commit.

```
1 http://quit.local/sparql
2 http://quit.local/sparql/<branchname>
3 http://quit.local/sparql/<commitId>
```

Listing 1: Dataset endpoint URLs.

*Git Interface.* The major operations on a versioning system are *commit* to create a new version in the version log as well as *merge*, and *revert* as operations on the version log. The commit operation is implemented by Update-Queries though the dataset access interface. To also allow the execution of *branch*, *merge*, and *revert* we make the operations provided by Git available through a web interface (cf. listing 2).

```
http://quit.local...
  /branch/<oldbranch>:<newbranch>
  /merge/<branch>:<target>?method=<strategy>
  /revert/<target>?commit=<commitId>
```

Listing 2: Merge and Revert Interfaces.

*Synchronization Between Instances.* So far all operations were executed on a local versioning graph that can diverge and be merged, but still, no collaboration with remote participants is possible. To allow collaboration on the World Wide Web, the versioning graph can be published (*push*) to a remote repository, from where other collaborators can copy (*clone*) the whole graph. If a collaborator already has cloned a previous version of the versioning graph, she can update here local graph by executing a *pull*. The default Git operations *pull* and *push* are also available via the Quit Store HTTP interface as shown in listing 3. They allow the user to synchronize with remote repositories.

```
http://quit.local/push/<remote name>/<local>:<remote branch>
http://quit.local/fetch/<remote name>/<remote branch>
http://quit.local/pull/<remote name>/<remote branch>:<local>
```

Listing 3: Push and Pull Interfaces.

[44] http://libgit2.github.com/
[45] http://www.pygit2.org/

### 9.2. Storing Provenance

Our system is built as a tool stack on top of Git to extend it with semantic capabilities. We focus on a generic solution to present provenance information that can be applied to arbitrary domains. Since our approach gains all of its versioning and storage capabilities from the underlying Git repository, most of the metadata is already captured and maintained (cf. *Use* and *Management*, [25]). With respect to the *Quit Stack* the two main concerns are, (1) to make the already existing provenance information from Git semantically available and (2) check how and to which extend additional and domain-specific metadata can be stored in the version control structure.

Our effort is to transform the metadata, stored in the data model of Git (cf. section 5.1) to RDF making use of PROV-O. Figure 10 provides an overview of provenance data that we provide for a commit. The *Git Commit*-sub-graph shows the information that could be extracted from the metadata associated with a commit in Git. Commits in Git can be mapped to instances of the class `prov:Activity` associated with their author and committer. We follow the idea of De Nies et al. [15] and represent the start and end time of the activity with the author and commit date of Git, which can be interpreted as the time till a change was accepted. PROV-O has no concept for commenting on activities, therefore we follow the suggestion of PROV-O and use `rdfs:comment` for commit messages. Git users are instances of `prov:Agent`, stored with their provided name and email. We represent Git names as `rdfs:label` since they do not necessarily contain a full name nor necessarily a nick name. Additionally the role of the user is provided through a `prov:Association`. To represent the roles used by Git we provide `quit:author` and `quit:Committer`. Further we store the commit ID using the custom property `quit:hex`.

In addition to the information that we can extract from the Git data model, we enrich the commits stored in Git with additional metadata. The two additional provenance operations that we support are *Import* and *Transformation*, which are shown on the left side in fig. 10. For an import we store `quit:Import`, a subclass of `prov:Activity`, together with a `quit:dataSource` property. The *Import* sub-graph contains the source specification of a dataset by recording its original URL on the Web. The *Transformation* sub-graph describes a change on the dataset that is represented as a `quit:Transformation` activity and a `quit:query` property to record the SPARQL Update Query, which was executed and resulted in the new commit. To enable the highest portability of this additional provenance data we want to persist it alongside Git's commit data structure. The main problem here is that Git itself offers no built-in feature to store any user-defined metadata for commits or files. What Git offers instead is a functionality called `git notes`, which is a commentary function on commits. Hereby, Git creates a private branch where text files, named after the commit they comment on, resides. The problem with this functionality is, that notes are not integrated into the commit storage. These notes are not included in the calculation of the object hash resp. commit ID and thus they are not protected against unperceived changes. Because we want to rely on all provenance information equally our decision is to provide and obtain additional metadata as part of the commit message as shown in listing 4. Commit messages are unstructured data, meaning it will not break the commit when additional structured data is provided at the start or end of a message. More specific key words to identify provenance operations other than *Source* and *Update* can be added as needed.

```
tree 31159f4524edf41e306c3c5148ed7734db1e777d
parent 3fe8fd20a44b1737e18872ba8a049641f52fb9ef
author pnaumann <patrick.naumann@stud.htwk-leipzig.de> ↵
  1487675007 +0100
committer pnaumann <patrick.naumann@stud.htwk-leipzig.de> ↵
  1487675007 +0100

Source: http://dbpedia.org/data/Leipzig.n3

Example Import
```

Listing 4: Git commit with additional data.

Additionally to the metadata that we can extract from a Git commit or that we encode in the commit message we have extended the graph with more information on the content of the dataset. In this case we store every named graph contained in a file as an instance of `prov:Entity`. We link the `prov:Entity` with the respective commit via `prov:wasGeneratedBy`. Each entity is attributed to the general graph URI using the `prov:specializationOf` property. The general graph is the URI of the graph under version control. Optionally we can also extend the provenance graph with the complete update information. For this purpose the `quit:update` property is used to reference the update node. The update node is linked to the original graph and the two named-graphs containing the added and deleted statements.

### 9.3. Accessing Provenance Information

To access our provenance information we follow the recommendation of the *PROV-AQ: Provenance Access and Query* (W3C Working Group Note)[46]. We provide two kinds of SPARQL interfaces, one service for the provenance graph and for each state in the history of the dataset an individual interface is provided (cf. section 9.1). The provenance graph is built from the metadata provided by Git and combined with the additional metadata stored in the commit messages. To be able to query this information we have to transform it to RDF and store the resulting graph. This is done during the initial start-up of the store system. The provenance graph is built from the commits stored in Git, by traversing the Git commit history of every branch from its end. The depth of the synchronized history as well as the selection of the relevant branches are configurable according to the users' needs. Therefore the needed storage space can be reduced for devices with low storage capacities, at the cost of time to parse graphs on-the-fly later on.

[46]https://www.w3.org/TR/prov-aq/

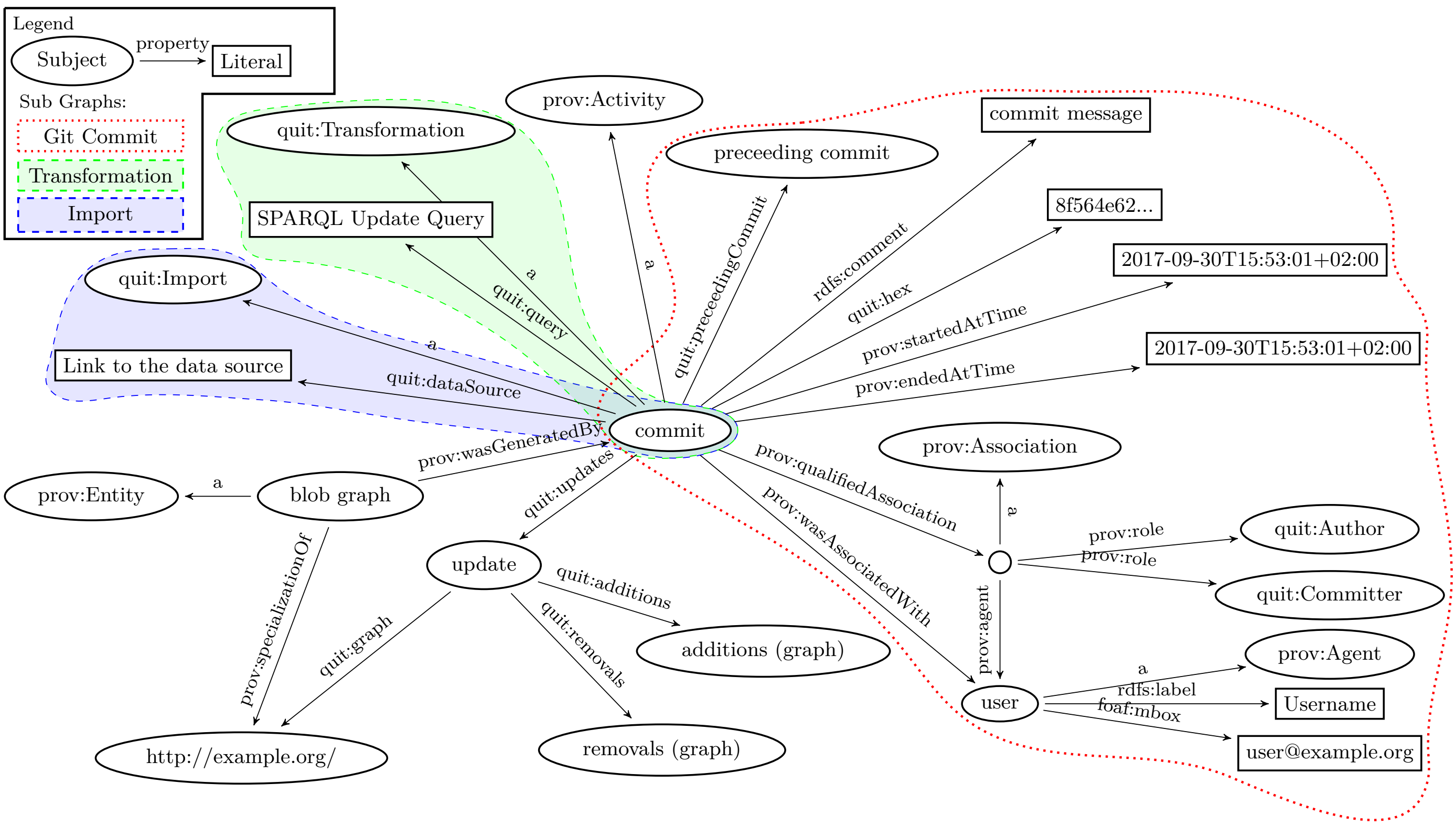

Figure 10: The provenance graph of a commit.

*Quit Blame.* As an example for the usage of provenance, similar to the functionality of `git blame`, we have also built a method to retrieve the origin of each individual statement in a dataset and associate it with its entry in the provenance graph. This is especially relevant for the aspects accountability and debugging as part of the aspect *Use* as described by [25]. Given an initial commit, we traverse the Git history to find the actual commit of each statement when it was inserted and annotate it with the metadata for that commit.

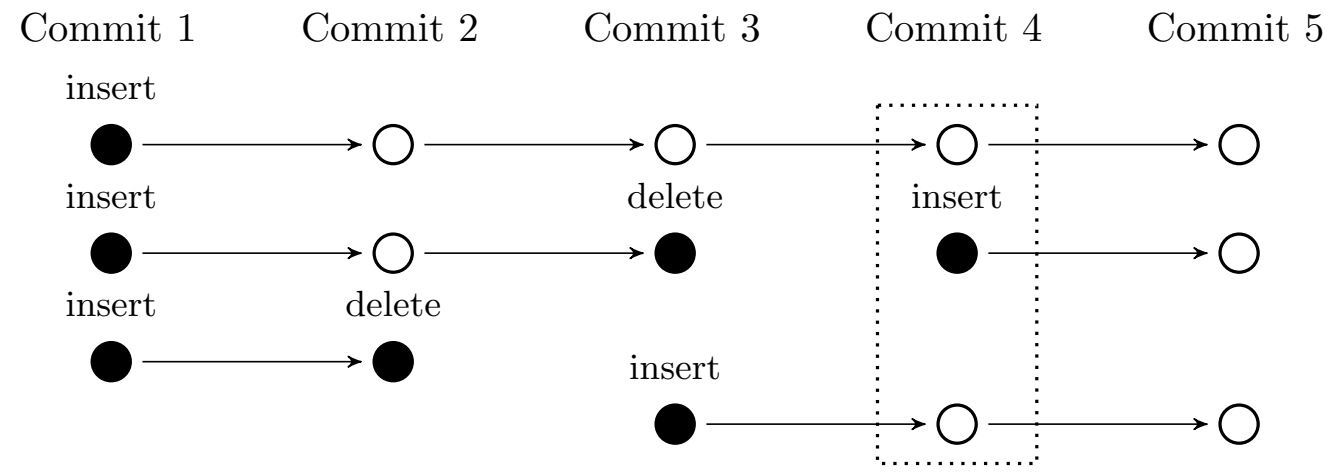

Figure 11: Example for an insert/delete chain in Git used by `git-blame`.

Looking at the example in fig. 11. The three statements that exist in the fourth commit should be matched with the commits 1, 4, and 3 respectively, since those are the commits where the statements were introduced. To implement this behavior on our provenance graph we utilize the SPARQL query as depicted in listing 5. As an input to the query we list all statements for which we want to identify the origin with subject `?s`, predicate `?p`, object `?o`, and named-graph `?context` in listing 5, line 13. For the execution of the query we loop through the list of commits starting at the current commit that is bound to the variable `?commit`. The query is then executed on the provenance graph until for all statements an originating commit was found.

```
1  SELECT ?s ?p ?o ?context ?commit ?name ?date WHERE {
2    ?commit prov:endedAtTime ?date ;
3            prov:wasAssociatedWith ?user ;
4            quit:updates ?update .
5    ?user   foaf:mbox ?email ;
6            rdfs:label ?name .
7    ?update quit:graph ?context ;
8            quit:additions ?additions .
9    GRAPH ?additions {
10     ?s ?p ?o
11   }
12   VALUES (?s ?p ?o ?context) {
13     ...
14   }
15 }
```

Listing 5: Query for `git blame` implementation.

## 10. Evaluation and Limitations

To evaluate the proposed framework we consider the correctness of the framework regarding the recorded changes and the performance, memory, and storage footprint. In order to pursue this task we have taken our implementation of the *Quit Store*[40]. This currently is a prototypical implementation to prove the concept of our framework, thus we are not aiming at competitive performance results. The hardware setup of the machine running the benchmarks is a virtual machine on a Hyper-V cluster with Intel(R) Xeon(R) CPU E5-2650 v3 with a maximum frequency of 2.30$GHz$ and 62.9$GiB$ of main memory. As operating system Ubuntu 16.10 (yakkety) 64-Bit is used.

```
$ ./generate -pc 4000 -ud -tc 4000 -ppt 1
```

Listing 6: The BSBM generate command with its argument.

As benchmarking framework we have decided to use the Berlin SPARQL benchmark (BSBM) [10], since it is made to execute SPARQL Query and SPARQL Update operations. The initial dataset as it is generated using the BSBM, shown in listing 6, contains 46370 statements and 1379201 statements to be added and removed during the benchmark. To also execute update operations we use the *Explore and Update Use Case*. We have executed 40 warm-up and 1500 query mix runs that resulted in 4592 commits on the underlying git repository using the testdriver as shown in listing 7.

```
$ ./testdriver http://localhost:5000/sparql \
  -runs 1500 -w 40 -dg "urn:bsbm" -o run.xml \
  -ucf usecases/exploreAndUpdate/sparql.txt \
  -udataset dataset_update.nt \
  -u http://localhost:5000/sparql
```

Listing 7: The BSBM testdriver command with its argument.

The setup to reproduce the evaluation is also available at the following link: `https://github.com/AKSW/QuitEval`.

### 10.1. Correctness of Version Tracking

To check the correctness of the recorded changes in the underlying git repository we have created a verification setup. The verification setup takes the git repository, the initial dataset, the query execution log (`run.log`) produced by the BSBM setup, and a reference store. The repository is set to its initial commit, while the reference store is initialized with the initial dataset. Each update query in the execution log is applied to the reference store. When an effect, change in number of statements[47], is detected on the reference store, the git repository is forwarded to the next commit. Now the content of the reference store is serialized and compared statement by statement to the content of the git repository at this point in time. This scenario is implemented in the `verify.py` script in the evaluation tool collection. We have executed this scenario and could ensure, that the recorded repository has the same data as the store after executing the same queries.

### 10.2. Correctness of the Merge Method

The functional correctness of the three-way merge method (cf. section 8.3) was verified using a repository filled with data using the graph generated by the BSBM. Since the three-way merge is conflict free it can be evaluated in a automated manner. To create the evaluation setup and run the verification of the results, a script was created. This script takes a git repository and creates a setup of three commits. An initial commit contains a graph file that serves as base of two branches. Each of the branches is forked from this initial commit and contains an altered graph file. The files in the individual commits contain random combinations of added and removed statements, while the script also produces the graph that is expected after merging the branches.

After creating the two branches with different graphs they are merged using `git merge`. The result is then compared statement by statement to the expected graph and the result is presented to the user. We have executed the verification 1000 times and no merge conflict or failure in the merge result occurred.

### 10.3. Context Merge

To demonstrate and evaluate the conflict identification of the *Context Merge*, we have created a repository that contains two branches holding the data as depicted in fig. 7. Further it contains a second graph with a resource with a misspelled label that is corrected on the *master* branch, while the resource is moved from the `http://example.org/…` namespace to `http://aksw.org/…` in the *develop* branch. The output of the context merge method is depicted in the screenshot fig. 12.

### 10.4. Query Throughput

The query throughput performance of the reference implementation was analyzed in order to identify obstacles in the conception of our approach. In fig. 13 the queries per second for the different categories of queries in the BSBM are given and compared to the baseline. Our baseline is the *Ad-hoc light weight SPARQL endpoint* (adhs)[48], a simple RDF store implemented using the Python RDFlib and a SPARQL interface implemented using Flask. To add support for quads we have adapted *adhs* accordingly, the source code can be found on GitHub[49]. We compare the execution of our store with version tracking enabled and with additionally enabled delta compression to the baseline *adhs*, which provides no version tracking. As expected the versioning has a big impact on the update queries (INSERT DATA and DELETE WHERE), while the explore queries (SELECT and CONSTRUCT) are not further impacted. We could reach $247 QMpH$[50] with Quit's versioning (*quit*), $235 QMpH$ with additionally enabled delta compression (cf. section 10.5) and $641 QMpH$ for the baseline. This is an improvement of $3.7\times$ over the speed of the old implementation of $67 QMpH$ (*old quit*), at the state of development as published in our previous paper [2].

[47] Since the Quit Store only creates commits for effective changes it is necessary to identify and skip queries without effect. This heuristic does not effect the result of the actual comparison, because still all queries are executed on the reference store, to which the content of the Quit Store is compared in each step.

[48] `https://github.com/nareike/adhs`

[49] `https://github.com/splattater/adhs/tree/feature/WSGI+QUADS`

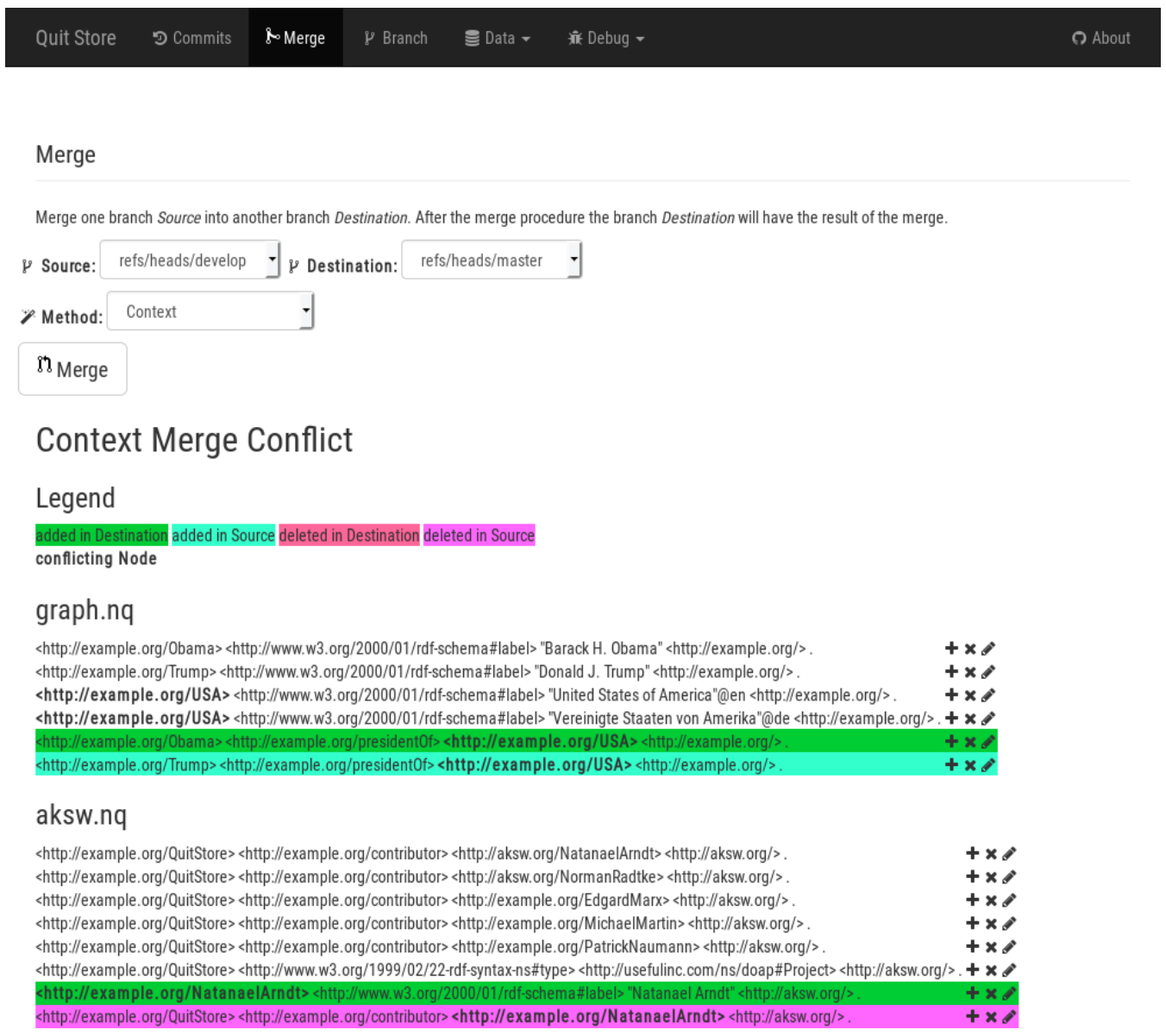

Figure 12: A merge on the quit store using the context strategy with identified conflicts.

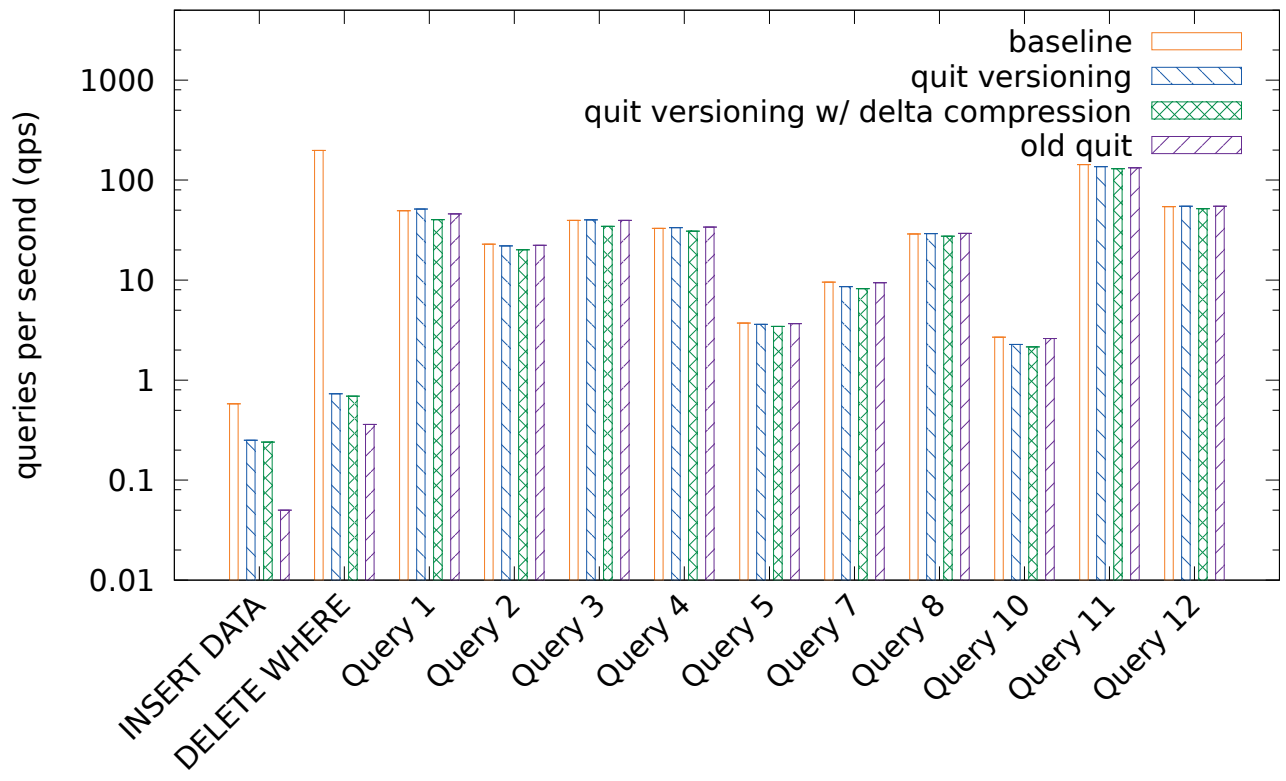

Figure 13: Execution of the different BSBM queries.

### *10.5. Storage Consumption*

During the executions of the BSBM we have monitored the impact of the Quit Store repository on the storage system. The impact on the storage system is visualized in fig. 14. We have measured the size of the repository on the left y-axis, and the size of the graph as well as the number of added and deleted statements on the right y-axis and have put it in relation to the number of commits generated at that point in time on the x-axis. We compare the execution of the BSBM on a Quit Store without using Git's delta compression (garbage collection[51]) and with the delta compression enabled. The repository has increased from initially $12.4 MiB$ to a size of $92.6 GiB$ without delta compression at the end of the benchmark. The benchmark started with $46,370$ initial statements and grew finally to $1,196,420$ statements. During the commits between 12 and 776 statements were added to the graph, while between 1 and 10 statements were removed. Enabling the compression feature during the benchmark could compress the repository to $18.3 GiB$. This is a compression rate of 80.2% at the end (at commit 4593). During the run of the evaluation the compression rate fluctuates, starting of course with 0% it is between 76.9% at commit 4217 and 94.5% at commit 4218 near the end of the evaluation run.

[50] $QMpH$ Query Mixes per Hour, Query Mixes are defined by the BSBM

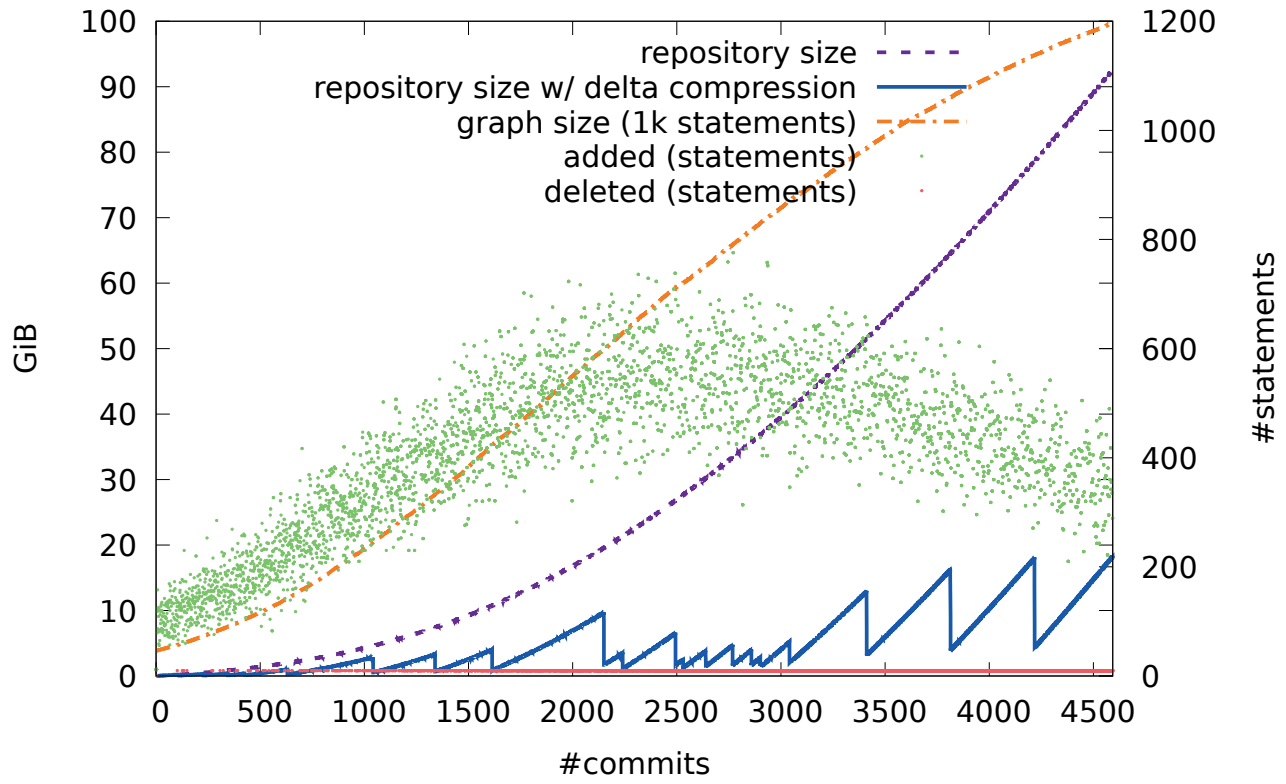

Figure 14: Storage consumption and number of commits in the repository during the execution of the BSBM.

### *10.6. Update Comparison*

To compare our prototypical implementation we have selected the implementations of the R43ples Store and the R&Wbase. To run the R&Wbase, the R43ples Store, and the Quit Store we have created Docker Images[52] for all systems that we have made available at the Docker Hub[53]. We have tried to execute the BSBM on the R43ples Store, but the queries were not actually executed and no revisions were created by the store during the run. We have also tried to execute the BSBM on the R&Wbase Store, but for each update query the generated version identifier of its predecessor is needed. This identifier has to be queried from the versioning graph before each new update query. This is not supported by the BSBM. Thus we have decided to create a custom comparison setup to measure the execution time of `INSERT DATA` and `DELETE DATA` queries. The comparison setup takes the `dataset_update.nt` file, which is generated by the BSBM, and randomly creates update queries. The system performs three steps, (1) generate a query log of `INSERT DATA` and `DELETE DATA` queries, (2) execute the generated queries on the store currently under testing, and (3) execute `SELECT` queries on a random sample of commits on the store under testing (cf. section 10.7). The system is available in our QuitEval repository (as mentioned above). On our system R&Wbase terminated after between 900 and 1900 commits for several query log configurations with the message: *Exception during abort (operation attempts to continue): Cannot abort a write lock-transaction*, when trying to query the versioning graph. Even after several attempts with exponential delay up to 1024 seconds, the system did not recover. We have found a setup that works for all stores. The setup creates update queries with 50000 total triples. Each generated update query has inserts or deletes a maximum of 200 statements. This setup results in 1370 update queries that produced 1370 versions on each store. The measured query execution times are depicted in figs. 15 to 17.

[51] `https://git-scm.com/docs/git-gc`

[52] Docker is a container system, see also `https://www.docker.com/`

[53] R43ples docker image: `https://hub.docker.com/r/aksw/r43ples/`, R&Wbase docker image: `https://hub.docker.com/r/aksw/rawbase/`, Quit Store docker image: `https://hub.docker.com/r/aksw/quitstore/`

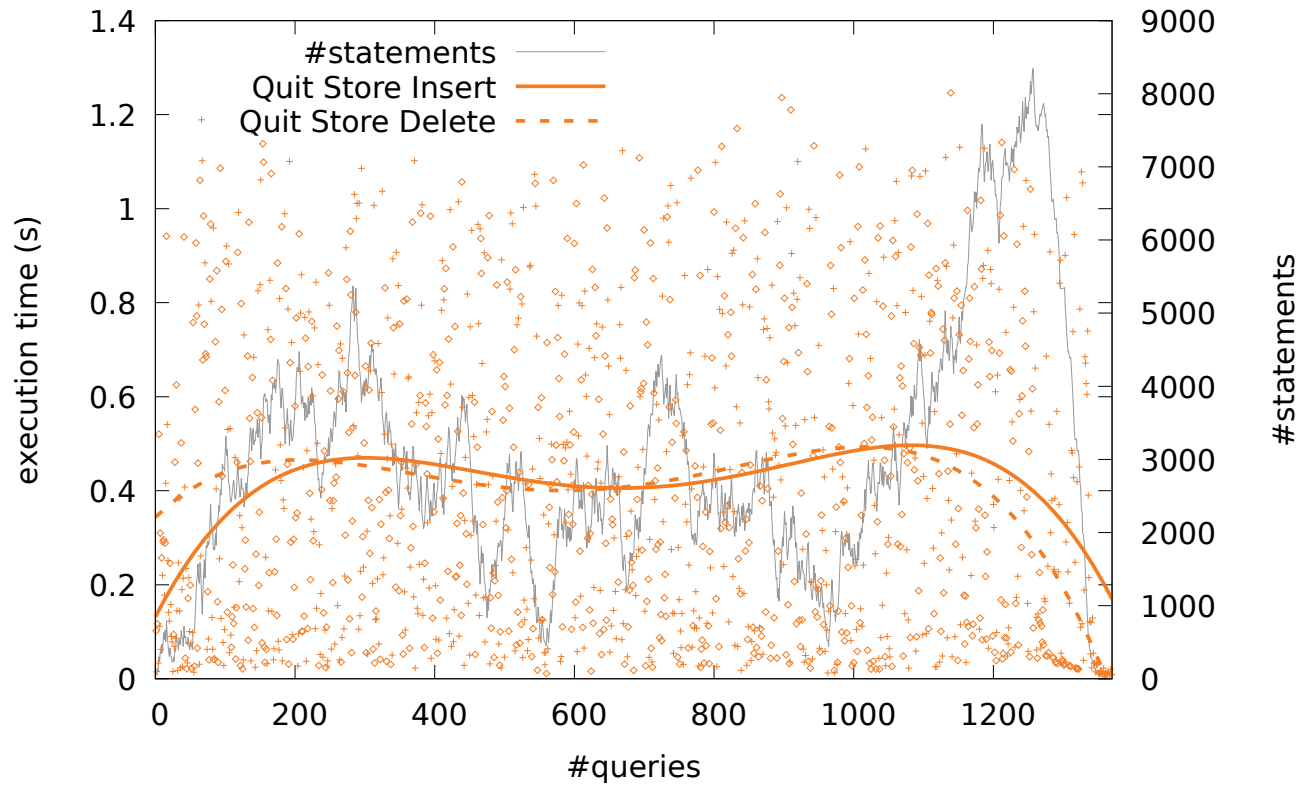

Figure 15: Query execution time of Quit Store with `INSERT DATA` and `DELETE DATA` queries and the number of statements (right y-axis).

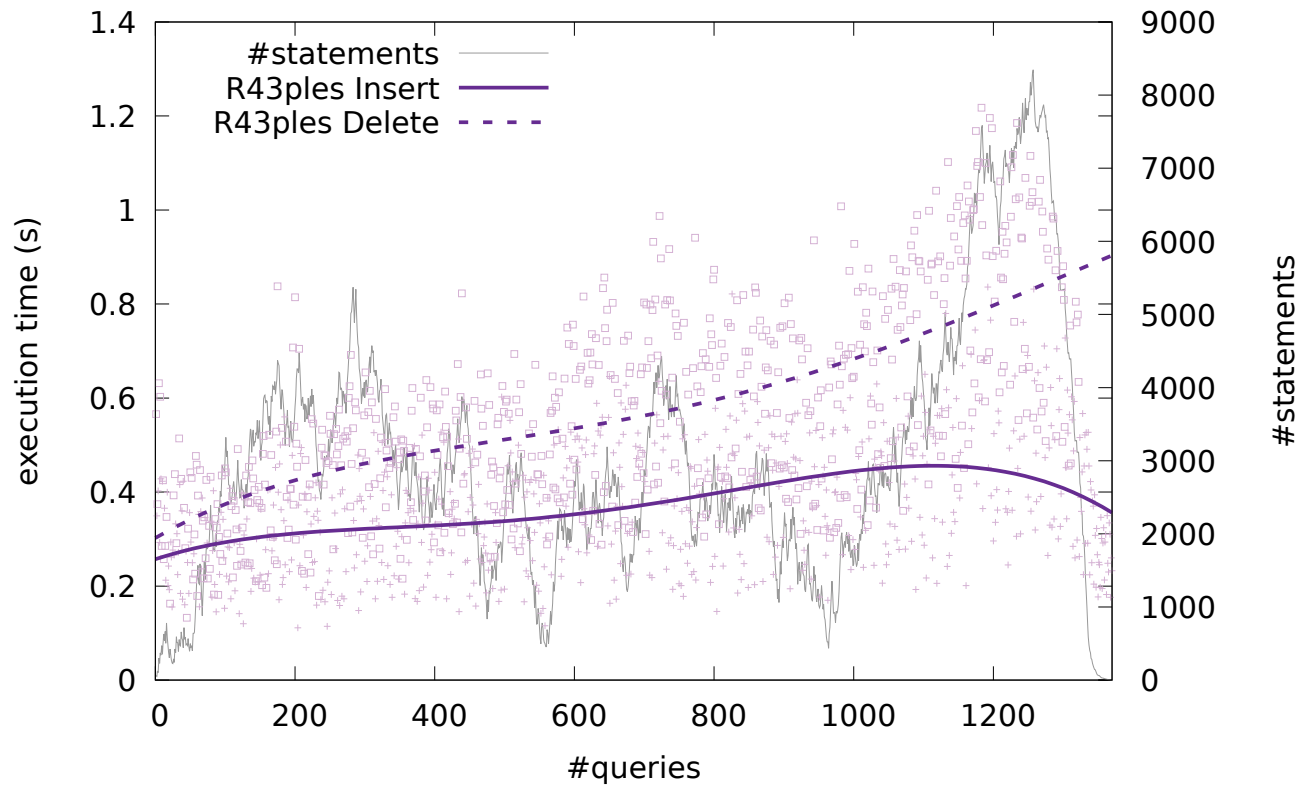

Figure 16: Query execution time of R43ples with `INSERT DATA` and `DELETE DATA` queries and the number of statements (right y-axis).

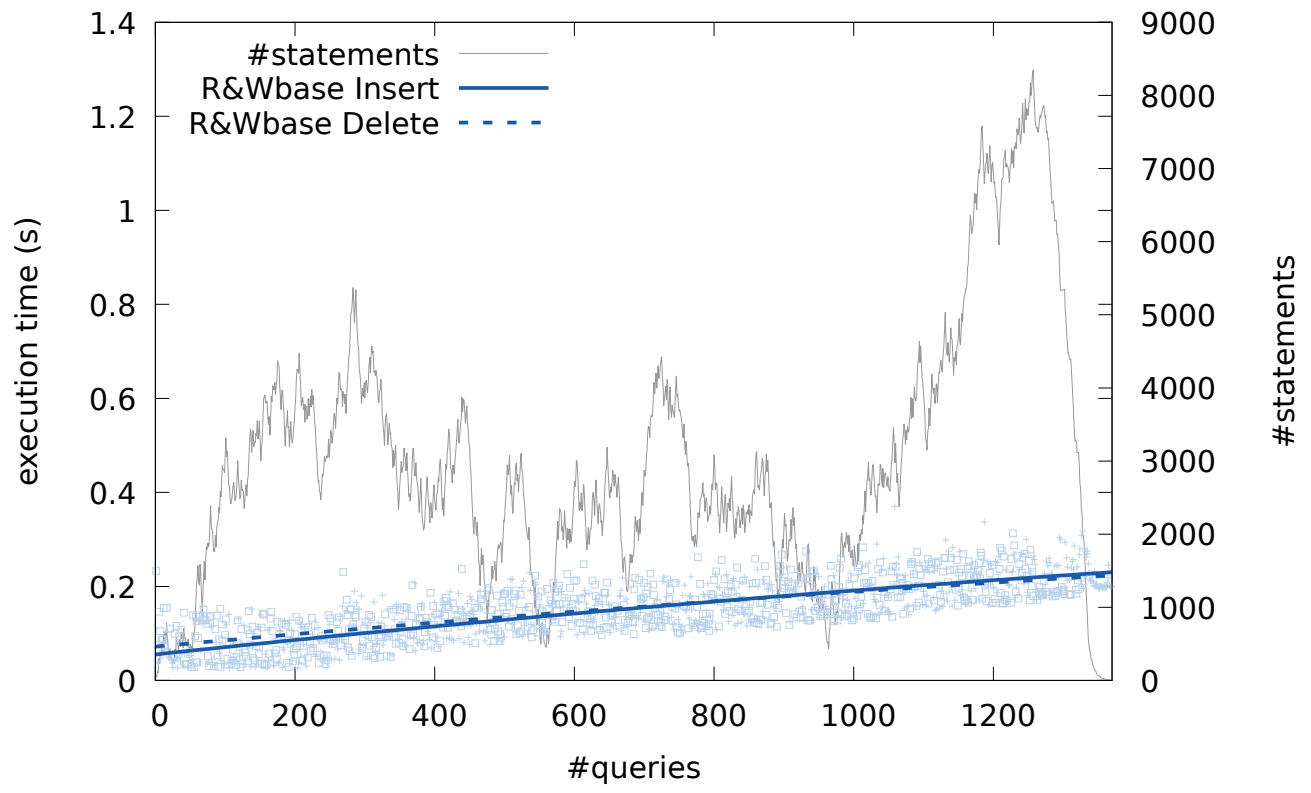

Figure 17: Query execution time of R&Wbase with `INSERT DATA` and `DELETE DATA` queries and the number of statements (right y-axis).

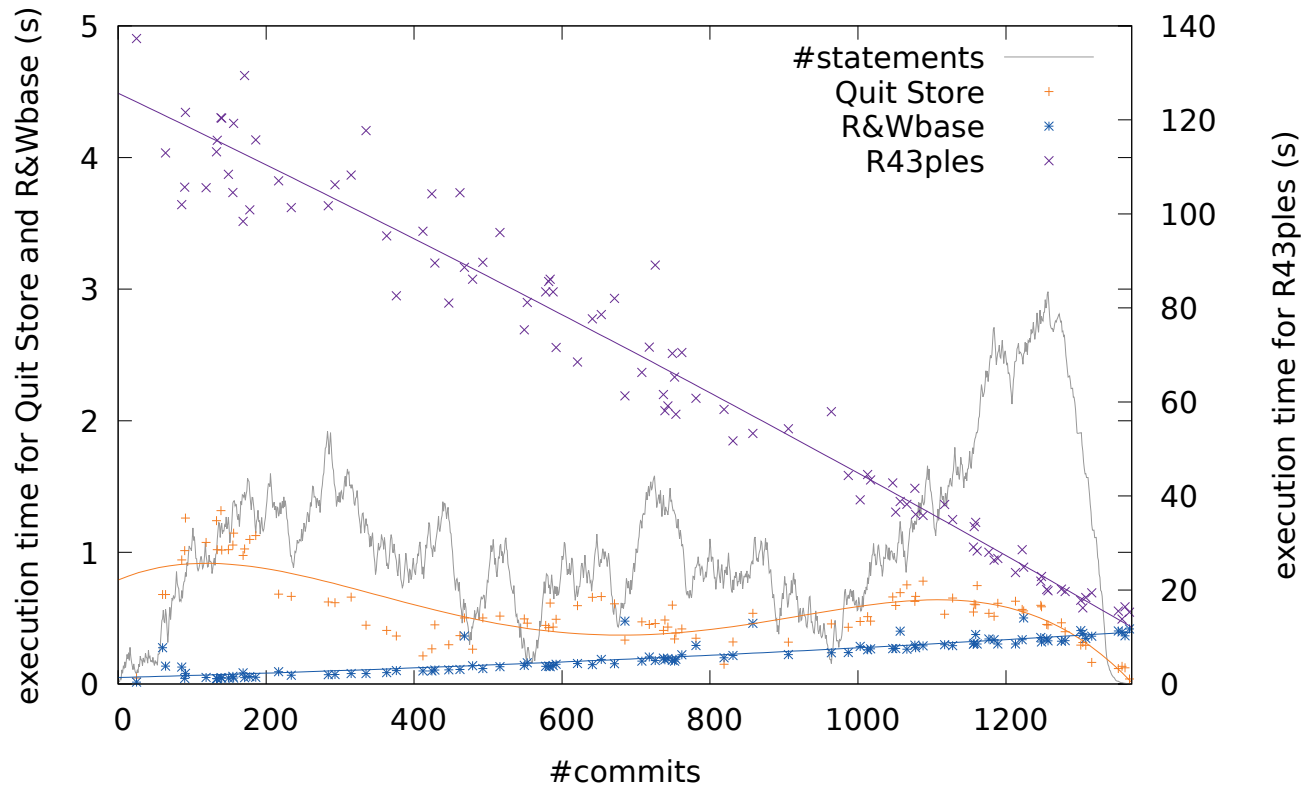

Figure 18: Query execution time comparison of Quit Store, R&Wbase, and R43ples (right y-axis) with `SELECT` queries on a random sample of revisions. (The number of revisions is just for illustration, for its scale please refer to figs. 15 to 17.)

### *10.7. Random Access*

After executing the update comparison we have also taken the resulting stores with all revisions and have performed a random access query execution comparison between R43ples, R&Wbase, and Quit Store. We have executed a simple select query (cf. listing 8) on a random sample of 100 of the recorded revisions on each store. The queries were executed in a random order on different revisions. The results of this comparison are depicted in fig. 18. The executions times of R43ples were significantly higher than the executions times of R&Wbase and Quit Store, we thus have decided to plot the results for R43ples on a different scale on the right y-axis.

```
SELECT ?s ?p ?o WHERE {
    GRAPH <urn:bsbm> {?s ?p ?o .}} LIMIT 1000
```

Listing 8: SELECT query used for the random access comparison.

## 11. Discussion

Based on the implementation of the Quit Store we have performed an evaluation regarding the correctness of the model (cf. sections 10.1 and 10.2) and monitored the performance of our implementation (cf. sections 10.4 to 10.7). With the improved implementation we could even reach a 3.6× improvement of the query throughput over the speed of the old implementation, at the state of development as published in our previous paper [2]. This improvement can mainly be led back to a completely reworked architecture, which is presented in section 9, and a number of optimizations based on profiling results, especially by reducing I/O overhead. The theoretical foundations could be confirmed. While the results show an acceptable performance the comparison to the base line should be an incentive for improvement.

The Quit Store is based on Git, which uses a snapshot based repository architecture. This snapshot storage was expected to increase severely as the stored dataset increases (cf. section 10.5), which could be confirmed during our evaluation. One of the disadvantages of this setup is also visible in fig. 14 towards the end of the evaluation run. Starting at about commit 3500, while the dataset slows down its growth, the repository is still growing by a copy of the changed graph plus the added statements. Here the delta compression comes into play. It could make up for the growth of the repository and we could show that the snapshot based approach does not necessarily put a high load on the storage requirements. Further we could show that the impact on the query execution performance was negligible (cf. section 10.4 and fig. 13). The evaluation of more advanced compression systems for the stored RDF dataset or the employment of a binary RDF serialization format, such as HDT [16] is still subject to future work.

To position our approach in relation to the related work we have compared the Quit Store with the R&Wbase and the R43ples implementations with respect to the execution time of `INSERT DATA` and `DELETE DATA` queries (cf. section 10.6 and figs. 15 to 17). We could show that the Quit Store's query execution time is related to the number of statements added resp. deleted. The R43ples Store's insert query execution times draw a similar picture as the Quit Store, because of its approach to keep the latest version as full snapshot in the revision log. In contrast the R43ples Store's delete query execution times and R&Wbase's insert and delete query execution times increase with the size of the repository. To also compare both stores with respect to the availability of the stored revisions we have performed random access select queries (cf. section 10.7 and fig. 18). With our random access query interface we can execute standard SPARQL queries on the store at the status of any revision at an overhead not related to the position of the revision in the repository. Due to the change based approach followed by R43ples and R&Wbase, these stores have high costs to retrieve old revisions (in case of R43ples), respective new revisions (in case of R&Wbase). Further the approach, as chosen for R43ples, of extending the syntax of standard SPARQL queries by custom keywords makes it harder to integrate the store with existing tools and systems. The necessity to query the versioning graph separately before creating a new version in R&Wbase, does also hinder its usability with standard tools.

We are able to track the provenance on any update operation in the dataset (UC 5 and REQ 4). With the provenance graph, we are also able to explore the recorded data using a standard SPARQL interface and due to its graph structure we are also able to represent any kind of a branched and merged history. Using `quit blame` we are able to track down the origin of any individual statement in a dataset. Due to the atomic level of provenance tracking the provenance information can be derived in two dimensions. From changes to individual atomic graphs we can directly conclude to a resource, graph, and dataset. In this way statements like "The dataset was changed by X on date Y" and also "The resource Z was changed by X on date Y" are possible. Also on the granularity level of the operations, multiple individual addition or deletion of an atomic graph are stored in a common commit alongside the executed query. Thus the provenance information is available on all levels of both dimensions *data-granularity* and *granularity of the change operation*.

With the context merge strategy we are able to identify possible conflicts and bring them to the users attention. This was demonstrated in section 10.3 with an example of two branches with conflicts in two graphs of the dataset. But the system can also be extended by custom merge tools. Besides the identification of conflicts during a merge process users can also define rules specific to the application domain, for instance by using ontological restrictions or rules (for instance using SHACL[54]). Support to ensure adherence to special semantic constraints, application specific data models, and certain levels of data quality is provided by continuous integration systems on the Git repository as presented in [31, 32, 36]. Employing continuous integration is already widely used in the Git ecosystem and can now also be adapted to RDF knowledge bases.

In the digital humanities projects *Pfarrerbuch*, with research communities in Hungary, Saxony Anhalt, and Saxony, as well as in the *Catalogus Professorum* project, the management and support of the diversity across its different datasets were made easy thanks to the adoption of *Quit*. The use of *Quit* allows different researchers to develop their datasets independently while sharing core components. For shared datasets it is now also possible to merge changes when there is a consensus to do so. This allows the digitalization team to continuously work in the data extraction and semantification while, the team of data curators can explore the extracted data and perform changes on the data. *Quit* also made it easy to explore differences across the different dataset versions by using the `diff` feature that previously had to be performed manually. Further it was possible to detect issues regarding the incorrect use of name spaces during the conversion process of the Hungarian Pastors dataset by using the provenance functionality. As a result it was also possible to solve this issue, by reverting the respective changes on the dataset and to deploy the updated dataset version.

[54] `https://www.w3.org/TR/shacl/`

## 12. Conclusion

Learning from software engineering history we could successfully adopt the distributed version control system *Git*. With the presented system we are now having a generic tool to support distributed teams in collaborating on RDF datasets. By supporting *commit* it is possible to track changes, made to a dataset, by *branching* the evolution of a dataset different points of view can be expressed (cf. REQ 1). Diverged branches can be consolidated using the *merge* operation while the user can select between different merge strategies (cf. REQ 2). Using the push and pull operations of *Quit* different instances can synchronize their changes and collaborate in a distributed setup of systems (cf. REQ 3).

The Quit Store tool provides a SPARQL 1.1 read/write interface to query and change an RDF dataset in a quad store that is part of a network of distributed RDF data repositories (cf. REQ 8). The store can manage a dataset with multiple graphs (cf. REQ 7), individual versions can be randomly selected from the repository (cf. REQ 5), and individual versions of the dataset can be compared using Quit Diff (cf. REQ 6). Based on the presented approach the application in distributed and collaborative data curation scenarios is now possible. It enables the setup of platforms similar to GitHub, specialized to the needs of data scientists and data engineers. It allows to create datasets by using local working copies and to send pull requests, while automatically keeping track of the data's versioning and provenance.

We have examined, how metadata and datasets stored in a Git repository can be enriched, processed, and used semantically. We have added methodologies for how Git commits, their metadata, and datasets can be used to store and exploit provenance information (cf. REQ 4). We could show that the concept of `git blame` can be transferred to semantic data using our provenance graph. With the presented system we can provide access to the automatically tracked provenance information with Semantic Web technology in a distributed collaborative environment.

In the future, *Quit* can support the application of RDF in enterprise scenarios such as supply chain management as described by Frommhold et al. [20]. An integration with the decentralized evolution model of distributed semantic social networks [6, 42], as well as the use case of synchronization in mobile scenarios [43] is possible. Further, we plan to lift the collaborative curation and annotation in distributed scenarios such as presented in the Structured Feedback protocol [1] to the next level by directly recording the user feedback as commits in a Quit Store. This setup can enable complex distributed collaboration strategies. As there is a big ecosystem of methodologies and tools around Git to support the software development process, the Quit Store can support the creation of such an ecosystem for RDF dataset management.

## 13. Acknowledgements

We want to thank the editors for organizing this special issue, our shepherd Olaf Hartig, and the reviewers for the critical and helpful reviews. Also we want to thank Sören Auer, Rafael Arndt, and Claudius Henrichs for their valuable remarks, important questions and for supporting us in proofreading. This work was partly supported by a grant from the German Federal Ministry of Education and Research (BMBF) for the LEDS Project under grant agreement No 03WKCG11C and the DFG project *Professorial Career Patterns of the Early Modern History: Development of a scientific method for research on online available and distributed research databases of academic history* under the grant agreement No 317044652.